\documentclass[twocolumn]{aastex63}

\usepackage{xspace}
\usepackage{epstopdf}
\usepackage{mathtools}
\usepackage{mathrsfs}
\usepackage{calrsfs}
\usepackage[hang,flushmargin]{footmisc}
\usepackage[autostyle, english = american]{csquotes}
\usepackage{multirow}
\MakeOuterQuote{"}

\received{--}
\revised{--}
\accepted{--}

\submitjournal{ApJ}

\shorttitle{A revised view of the linear polarization in the subparsec core of M87 at 7 mm}
\shortauthors{Park et al.}

\graphicspath{{./}}

\begin{document}

\title{A revised view of the linear polarization in the subparsec core of M87 at 7 mm}

\correspondingauthor{Jongho Park}
\email{jpark@asiaa.sinica.edu.tw}

\author[0000-0001-6558-9053]{Jongho Park}
\affiliation{Institute of Astronomy and Astrophysics, Academia Sinica, P.O. Box 23-141, Taipei 10617, Taiwan}

\author[0000-0001-6988-8763]{Keiichi Asada}
\affiliation{Institute of Astronomy and Astrophysics, Academia Sinica, P.O. Box 23-141, Taipei 10617, Taiwan}

\author[0000-0001-6081-2420]{Masanori Nakamura}
\affiliation{National Institute of Technology, Hachinohe College, 16-1 Uwanotai, Tamonoki, Hachinohe, Aomori 039-1192, Japan}
\affiliation{Institute of Astronomy and Astrophysics, Academia Sinica, P.O. Box 23-141, Taipei 10617, Taiwan}

\author[0000-0002-2709-7338]{Motoki Kino}
\affiliation{Kogakuin University of Technology \& Engineering, Academic Support Center, 2665-1 Nakano, Hachioji, Tokyo 192-0015, Japan}
\affiliation{National Astronomical Observatory of Japan, 2-21-1 Osawa, Mitaka, Tokyo 181-8588, Japan}

\author[0000-0001-9270-8812]{Hung-Yi Pu}
\affiliation{Department of Physics, National Taiwan Normal University, No. 88, Sec. 4, Tingzhou Rd., Taipei 116, Taiwan, R.O.C.}
\affiliation{Institute of Astronomy and Astrophysics, Academia Sinica, P.O. Box 23-141, Taipei 10617, Taiwan}

\author[0000-0001-6906-772X]{Kazuhiro Hada}
\affiliation{Mizusawa VLBI Observatory, National Astronomical Observatory of Japan, Osawa, Mitaka, Tokyo 181-8588, Japan}
\affiliation{Department of Astronomical Science, The Graduate University for Advanced Studies (SOKENDAI), 2-21-1 Osawa, Mitaka, Tokyo 181-8588, Japan}

\author[0000-0002-2709-7338]{Evgeniya V. Kravchenko}
\affiliation{Moscow Institute of Physics and Technology, Institutsky per. 9, Moscow region, Dolgoprudny, 141700, Russia}
\affiliation{Astro Space Center, Lebedev Physical Institute, Profsouznaya 84/32, Moscow 117997, Russia}

\author[0000-0002-8657-8852]{Marcello Giroletti}
\affiliation{INAF Istituto di Radioastronomia, Via P. Gobetti, 101, I-40129 Bologna, Italy}

\begin{abstract}

The linear polarization images of the jet in the giant elliptical galaxy M87 have previously been observed with Very Long Baseline Array at 7 mm. They exhibit a complex polarization structure surrounding the optically thick and compact subparsec-scale core. However, given the low level of linear polarization in the core, it is required to verify that this complex structure does not originate from residual instrumental polarization signals in the data. We have performed a new analysis of the same data sets observed in four epochs by using the Generalized Polarization CALibration pipeline (GPCAL). This novel instrumental polarization calibration pipeline overcomes the limitations of LPCAL, a conventional calibration tool used in the previous M87 studies. The resulting images show a compact linear polarization structure with its peak nearly coincident with the total intensity peak, which is significantly different from the results of previous studies. The core linear polarization is characterized as fractional polarization of $\sim0.2$--0.6\% and polarization angles of $\sim66$--$92^\circ$, showing moderate variability. We demonstrate that, based on tests with synthetic data sets, LPCAL using calibrators having complex polarization structures cannot achieve sufficient calibration accuracy to obtain the true polarization image of M87 due to a breakdown of the "similarity approximation". We find that GPCAL obtains more accurate D-terms than LPCAL by using observed closure traces of calibrators that are insensitive to both antenna gain and polarization leakage corruptions. This study suggests that polarization imaging of very weakly polarized sources has become possible with the advanced instrumental polarization calibration techniques.

\end{abstract}


\section{Introduction} 
\label{sec:intro}

Collimated outflows, called jets, from active galactic nuclei (AGNs) are among the most energetic phenomena in the Universe. They often move at relativistic speeds \citep[e.g.,][]{Jorstad2017, Lister2018}, produce high-energy photons at X-rays and $\gamma$-rays \citep[e.g.,][]{Kataoka2006, MS2016, Blandford2019}, and affect the interstellar and intergalactic medium by transferring momentum and energy \citep[e.g.,][]{Fabian2012, Yuan2018}. The magnetic field around supermassive black holes is thought to play a critical role in launching the jets \citep[e.g.,][]{BZ1977, BP1982, Narayan2003, NQ2005, McKinney2006, Tchekhovskoy2011}.

In April 2019, the Event Horizon Telescope collaboration (EHTC) reported a ring-like structure at 1.3 mm on event horizon scales of the supermassive black hole at the center of the giant elliptical galaxy M87 \citep{EHT2019a, EHT2019b, EHT2019c, EHT2019d, EHT2019e, EHT2019f}. Very recently, the EHTC has presented corresponding linear-polarimetric images of M87 \citep{EHT2021a}. Interestingly, the polarization position angles of the ring are arranged in a nearly azimuthal pattern. Detailed modeling of these images using general relativistic magnetohydrodynamic (GRMHD) simulations suggested that near-horizon magnetic fields are dynamically important \citep{EHT2021b}, which can produce powerful jets efficiently \citep[e.g.,][]{Tchekhovskoy2011, Sadowski2013}.

Magnetic fields are not only crucial for jet launching but also jet acceleration. MHD models predict that jets can be accelerated to relativistic speeds by transferring Poynting flux to kinetic energy flux through the so-called magnetic nozzle effect \citep[e.g.,][]{Li1992, BL1994, VK2004, Komissarov2007, Tchekhovskoy2008, Lyubarsky2009, PT2020}. During this process, systematic evolution of the magnetic field from poloidal to toroidal-dominated configurations is expected \citep[e.g.,][]{VK2004, McKinney2006, Komissarov2007, Komissarov2009, McKinney2012, Pu2015}. However, there is no clear observational evidence for this systematic transition yet. The main reason is that jet acceleration is expected to occur at distances less than $\approx10^4$--$10^6$ Schwarzschild radii ($R_S$) from the black hole \citep{VK2004, Marscher2008, Meier2012, Boccardi2016b, Mertens2016, Hada2017, Park2019b, Park2021b}), which can be well resolved for nearby radio galaxies only. Nearby radio galaxies, though, are usually unpolarized or very weakly polarized in the expected jet acceleration regions \citep[e.g.,][]{Nagai2017, Lister2018, Kim2019, Park2019a}, making it difficult to study the magnetic fields. 

M87 is currently the best laboratory for studying jet magnetic fields and MHD models. It hosts a supermassive black hole with a mass of $M_{\rm BH} = (6.5\pm0.7)\times10^9\ M_\odot$ (\citealt{EHT2019f}; see also \citealt{Gebhardt2011, Walsh2013}), is located at a distance of 16.8 Mpc \citep[][]{Blakeslee2009, Bird2010, Cantiello2018, EHT2019f}, giving a scale of $1\ {\rm mas} \approx131\ R_S$. M87 has a bright, straight jet extending northwest \citep[e.g.,][]{Owen1989, Biretta1995, Perlman1999, Meyer2013, EHTMWL2021} and a weak counterjet that can only be seen on subparsec scales \citep[e.g.,][]{Ly2007, Kovalev2007, Hada2018, Kim2018, Walker2018}. Previous very long baseline interferometric (VLBI) observations show that the jet is systematically collimated \citep[e.g.,][]{AN2012, Hada2013, NA2013} and accelerated to relativistic speeds \citep[][]{Asada2014, Mertens2016, Hada2017, Nakamura2018, Walker2018, Park2019b} at deprojected jet distances $\lesssim5\times10^5\ R_S$. Furthermore, \cite{Park2019a} showed that the magnitude of Faraday rotation measures (RMs) in the jet systematically decreases with increasing jet distance in the jet acceleration and collimation zone, which indicates that the magnetic field in the jet and its environment may evolve with distance from the black hole.

\cite{Walker2018} presented linear polarization images of M87 based on Very Long Baseline Array (VLBA) observations at 43 GHz in two epochs in 2007. The inferred magnetic field vectors, assumed to be perpendicular to the observed linear polarization vectors, wrap around the total intensity core. They interpret this result as a sign of a toroidal jet magnetic field geometry. \cite{Kravchenko2020} extended this study and presented similar linear polarization structures near the core based on more VLBA data sets at 22 and 43 GHz. These results indicate that toroidal magnetic fields may be already dominant in the jet on scales of a few hundreds $R_S$, which can put constraints on the MHD models.

However, this kind of polarimetric structure, to our knowledge, has not been observed in most other AGN jets, which usually show a compact polarization structure near the core \citep[e.g.,][]{Jorstad2017, Lister2018}. Some blazars do show complex polarization structures near the cores \citep[e.g.,][]{Cawthorne2013, Marscher2016}, which are believed to be associated with recollimation shocks formed in the jets at relatively large distances from the black hole (e.g., $\gtrsim10^4\ R_S$; \citealt{Marscher2008}). However, the subparsec-scale radio core in M87 at 43 GHz is expected to be located at a deprojected distance of $\approx18\ R_S$ from the black hole for the assumed black hole mass of $6.5\times10^9\ M_\odot$ \citep{EHT2019f} and the viewing angle of $17^\circ$ \citep{Walker2018}, based on the core-shift measurements by \cite{Hada2011}. Therefore, it may be possible that M87 exhibits, thanks to its proximity and large black hole mass, extraordinary linear polarization structures that have not been observed in other AGN jets. Nevertheless, since the observed linear polarization fraction is quite low ($\lesssim1.5\%$; \citealt{Walker2018}), one must verify that the images are not significantly affected by residual instrumental polarization signals in the data.

Typically, VLBI data has instrumental polarization signals due to systematic corruption caused by unwanted leakage of opposite-handed polarization signals to each feed. This polarization leakage (also known as "D-terms") must be removed from the data before producing source polarization images. Any residual leakage in data would produce spurious artificial polarization signals, whose polarization intensity scales with the total intensity of the source \citep[e.g.,][]{Roberts1994, Leppanen1995, Hovatta2012}. Therefore, bright cores in AGN jets can be significantly affected by residual polarimetric leakages in data, especially when the source fractional polarization is lower than or comparable to the level of residual instrumental polarization.

The measured cross-hand visibilities contain both source-intrinsic linear polarization signals and instrumental polarization signals (see Equation~\ref{eq:meas}). The former rotates with antenna field rotation angles\footnote{The field rotation angles are equivalent to the parallactic angles for alt-azimuth feeds \citep[see, e.g.,][]{Park2021a}.}, while the latter is independent of the field rotation angles (before applying field rotation angle corrections to the data). This difference makes it easier to disentangle the two signals. LPCAL \citep{Leppanen1995} is a task implemented in the Astronomical Image Processing System (AIPS; \citealt{Greisen2003}), which has been widely used for instrumental polarization calibration of VLBI data, including the previous M87 core polarization studies \citep{Walker2018, Kravchenko2020}. The source polarization signals are usually unknown and thus some assumptions are needed to derive D-terms. LPCAL divides the total intensity model of a calibrator into several sub-models and assumes that each sub-model has a constant fractional polarization and electric vector position angle (EVPA), so-called the "similarity approximation" \citep{Cotton1993, Leppanen1995}. Thus, LPCAL works well for deriving accurate D-terms if there are calibrators that are unpolarized or have compact structures.

However, this approximation may not always hold, especially for VLBI data at high frequencies in which nearly all calibrators are resolved. They often show the total intensity and linear polarization structures not similar to each other \citep[e.g.,][]{Jorstad2017, Lister2018}, resulting in a breakdown of the similarity approximation. To overcome the limitations of LPCAL, we have developed the Generalized Polarization CALibration pipeline (GPCAL; \citealt{Park2021a}), which is written in ParselTongue \citep{Kettenis2006} and based on AIPS and the Caltech Difmap package \citep{Shepherd1997}. With GPCAL, more accurate linear polarization models of calibrators can be used for D-term estimation without being limited by the similarity approximation. Also, it can fit the instrumental polarization model to data from multiple calibrator sources simultaneously to increase the fitting accuracy. Furthermore, LPCAL uses a linear approximation (i.e., ignoring the last terms in Equation~\ref{eq:meas}), while GPCAL does not. Thus, GPCAL deals with data having large D-terms (e.g., $\gtrsim10$\%) better than LPCAL. GPCAL has already been applied to the polarization analysis of the first M87 EHT results\footnote{Other techniques used for the EHT analysis are \texttt{polsolve} \citep{MartiVidal2021}, similar to GPCAL but based on CASA \citep{McMullin2007, Janssen2019}, the \texttt{eht-imaging} software library using the regularized maximum likelihood technique \citep{Chael2016, Chael2018}, D-term Modeling Code (DMC, \citealt{Pesce2021}) and THEMIS \citep{Broderick2020} using Markov chain Monte Carlo schemes.} \citep{EHT2021a}.

In this paper, we revisit the VLBA data sets presented in \cite{Walker2018} and \cite{Kravchenko2020}. In Section~\ref{sec:vlba}, we obtain new linear polarization images of M87 with GPCAL and LPCAL, and compare the results. In Section~\ref{sec:synthetic}, we test GPCAL and LPCAL by using synthetic data sets. We demonstrate that D-terms obtained with LPCAL using calibrators having complex linear polarization structures are not accurate enough to reconstruct the weak polarization structure of M87. In Section~\ref{sec:ctrace}, we evaluate the performance of different instrumental polarization calibration strategies by using the observed "closure traces", which are quantities insensitive to both antenna gain and leakage corruptions \citep{BP2020}. In Section~\ref{sec:discussion}, we compare the observed M87 core polarization at 43 GHz with previous observations at other frequencies and discuss the prospect for future quasi-simultaneous multifrequency VLBI observations. We summarize and conclude in Section~\ref{sec:conclusion}.

\section{VLBA archive data analysis} 
\label{sec:vlba}

\begin{deluxetable*}{cccccccc}
\tablecaption{Archival VLBA Data of M87\label{tab:data}}
\tablewidth{0pt}
\tablehead{\colhead{Proj. Code} & \colhead{Obs. Date} & \colhead{Antennas} & \colhead{Beam Shape} & \colhead{$I_{\rm core}$} & \colhead{$P_{\rm core}$} & \colhead{$m_{L, {\rm core}}$} & \colhead{$\chi_{\rm core}$} \\
&& & (mas $\times$ mas, degree) & \multicolumn{2}{c}{$\rm (mJy\ Beam^{-1})$} & (\%) & (degrees) \\
&& (a) & (b) & (c) & (d) & (e) & (f)
}
\startdata
\hline
BW088A & 2007 Jan 27 & VLBA, -KP & $0.40\times0.20, -5.3^\circ$ & $730\pm0.2$ & $4.21\pm0.28$ & $0.58\pm0.04$ & $91.8\pm3.55$ \\
BW088G & 2007 May 09 & VLBA & $0.42\times0.21, -7.8^\circ$ & $699\pm0.2$ & $2.26\pm0.22$ & $0.32\pm0.03$ & $69.9\pm4.07$ \\
BG250A & 2018 Apr 28 & VLBA, -HN, -PT & $0.42\times0.22, -6.6^\circ$ & $603\pm0.1$ & $1.44\pm0.15$ & $0.24\pm0.02$ & $91.4\pm3.54$ \\
BG250B1 & 2018 May 25 & VLBA, -OV & $0.40\times0.19, -6.5^\circ$ & $464\pm0.1$ & $1.17\pm0.14$ & $0.25\pm0.03$ & $66.3\pm3.94$ \\
\hline
\enddata
\tablecomments{(a) Participating antennas. VLBA means that all ten antennas are present in the data. The minus sign indicates that the data for that antenna is missing. (b) Major axis, minor axis, and position angle of the synthesized beam under the natural weighting of the data. (c) Total intensity of the core in units of mJy per beam. The uncertainties are calculated from the off-source image rms noise, which does not include the systematic errors caused by inaccurate antenna sensitivity measurements, pointing offsets, and so on. (d) Linear polarization intensity at the core. (e) Linear polarization fraction at the core in units of percent. (f) EVPA at the core in units of degree.}
\end{deluxetable*}

We analyzed four VLBA data sets at 43 GHz in the NRAO archive. We summarize the basic properties of the data sets in Table~\ref{tab:data}. The linear polarization images of M87 obtained from the two data sets in 2007 and all four data sets were presented in \cite{Walker2018} and \cite{Kravchenko2020}, respectively. We performed a standard data reduction with AIPS, as described in \cite{Park2021b}. There are two calibrators in the data sets: 3C 279 and OJ 287. We performed CLEAN and phase/amplitude self-calibration iteratively with Difmap and produced total intensity images of M87 and the calibrators.

We adopt two methods for deriving D-terms and compare the resulting polarization images of M87. One is using LPCAL on each calibrator, as in the previous studies \cite[][]{Walker2018, Kravchenko2020}. The other is using GPCAL on the data of all three sources (M87, 3C 279, and OJ 287). The two data sets observed in 2007 consist of two baseband channels (often called "IFs") with a bandwidth of 16 MHz per IF. The other two data sets observed in 2018 consist of eight IFs with a bandwidth of 32 MHz per IF. We solved for D-terms for different IFs independently for both methods and used the data averaged over frequency within each IF. 

The first method derives D-terms using LPCAL on individual calibrators. The total intensity images of the calibrators consist of several knots. We divided their CLEAN models into several submodels in such a way that each knot is regarded as a submodel. LPCAL assumes that the linear polarization structure of each submodel is proportional to its total intensity structure \citep{Leppanen1995}. We also derived D-terms using LPCAL on M87, assuming that it is unpolarized. This is a reasonable assumption as the M87 jet is unpolarized for most jet regions on VLBI scales \cite[e.g.,][]{ZT2002, Park2019a}.

The second method used a similar GPCAL pipeline to that used for the VLBA data calibration in \cite{Park2021a}. The pipeline first derives D-terms using M87, assuming that it is unpolarized. The D-terms are removed from the data of calibrators using this initial estimate. The pipeline then performs additional instrumental polarization self-calibration with 10 iterations using 3C 279 and OJ 287. The complex linear polarization structures of the calibrators are considered for D-term estimation at this stage, allowing us to take advantage of their high signal-to-noise ratios (S/Ns). The pipeline removes the D-terms from the M87 data using the final D-term estimates. We found that the results do not change significantly if we used the calibrators for the initial D-term estimation as well. The resulting linear polarization images of M87 from this calibration are presented in Appendix~\ref{appendix:noM87}.

We corrected the remaining phase offset between the right and left hand circular polarizations (RCP and LCP) at the reference antenna for each IF by comparing the integrated EVPAs of OJ 287 with contemporaneous (within two weeks of the VLBA observations) Very Large Array (VLA) observations\footnote{\url{http://www.aoc.nrao.edu/~smyers/calibration/}} for the BW088A and BW088G data sets. OJ 287 showed small magnitudes of Faraday rotation measures ($\rm |RM| \lesssim500{\ \rm rad\ m^{-2}}$), good $\lambda^2$ laws for EVPAs between 8 and 43 GHz, and stable EVPAs in the VLA data. In this case, additional uncertainty from the EVPA calibration is expected to be small, and we assume that the calibration is accurate within $\pm3^\circ$ \citep[e.g.,][]{OG2009}. We used contemporaneous Korean VLBI Network (KVN) single-dish observations of OJ 287 and 3C 279 at 43 GHz for the EVPA calibration of the other two VLBA data sets, part of a project named "PAGaN"\footnote{\url{https://radio.kasi.re.kr/kvn/ksp.php}.} that monitors bright blazars with the KVN at 22, 43, 86, and 129 GHz every month (see \citealt[][]{Park2018} for more details). We found that both 3C 279 and OJ 287 showed small RM magnitudes and stable EVPAs in the KVN data. The small RMs allow us to use the VLA data or the KVN single-dish data for the EVPA calibration of the VLBA data despite the slight difference in their observing frequencies. Comparing the EVPA correction factors derived from the two sources, we find that the calibration using the KVN data is accurate within $\approx2^\circ$. After the EVPA calibration, the final linear polarization images of M87 were produced using CLEAN in Difmap. 

\begin{figure*}[t!]
\centering
\includegraphics[width = 1.0\textwidth]{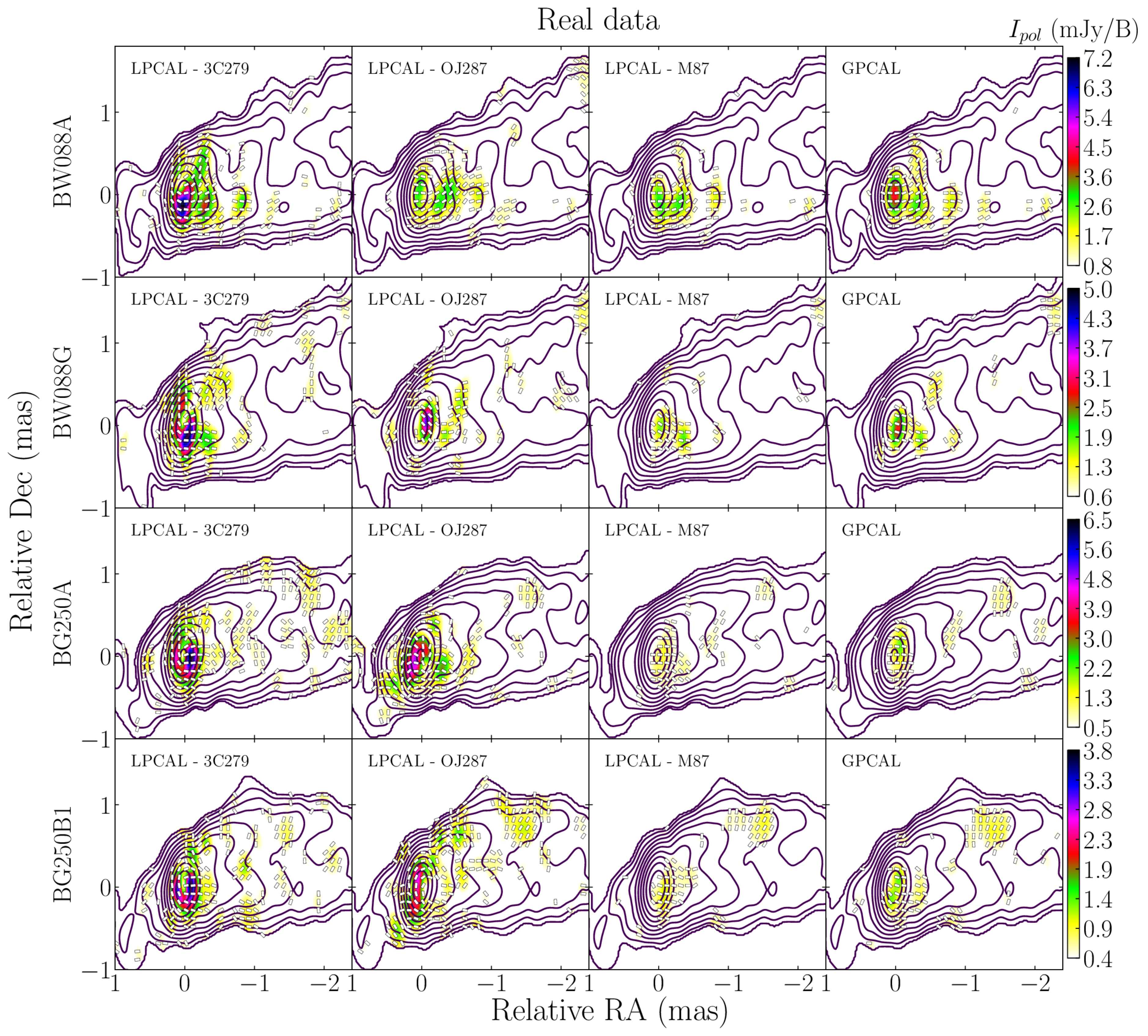}
\caption{Linear polarization images of M87 obtained from VLBA data sets observed in four epochs at 43 GHz. The results for each data are presented in each row. The polarization images of different instrumental polarization calibration versions, using LPCAL on individual sources (3C 279, OJ 287, and M87) and using GPCAL, are shown in different columns. Color shows the distribution of Ricean de-biased linearly polarized intensity, and the white ticks show EVPAs. \label{fig:realdata}}
\end{figure*}

We present the linear polarization images of M87 obtained from different versions of instrumental polarization calibration in Figure~\ref{fig:realdata}. We have made Ricean de-biasing corrections, i.e., $P_{\rm corr} = P_{\rm obs}\sqrt{1 - (\sigma_P/P_{\rm obs})^2}$ \citep{WK1974, Thompson2017}, where $P_{\rm corr}$ and $P_{\rm obs}$ denote the de-biased and observed polarized intensity, respectively, and $\sigma_P$ is the average of the off-source rms noise levels in Stokes $Q$ and $U$ maps \citep{Hovatta2012, Lister2018}. Interestingly, the resulting M87 polarization structure varies significantly from calibration version to version. The structures presented in the previous studies \citep[][]{Walker2018, Kravchenko2020} could be reproduced when using LPCAL on 3C 279\footnote{Both \cite{Walker2018} and \cite{Kravchenko2020} used LPCAL on 3C 279 for instrumental polarization calibration of the data sets presented in this paper (R. C. Walker, private communication; see Appendix A in \citealt{Kravchenko2020}).}. However, the complex polarization structures disappear when using LPCAL on the other sources (OJ 287 and M87) or using GPCAL. The polarization intensity peak location near the core is nearly coincident with the total intensity peak location when using LPCAL on M87 or using GPCAL, as observed in most other radio-loud AGNs \citep[e.g.,][]{Jorstad2017, Lister2018}. In Table~\ref{tab:data}, we provide the total intensity ($I_{\rm core}$), linearly polarized intensity ($P_{\rm core}$), fractional polarization ($m_{L, {\rm core}}$), and EVPA ($\chi_{\rm core}$) at the core of the images obtained with GPCAL. We identified the core position with the total intensity peak position in the image.

\cite{Kravchenko2020} presented the linear polarization image of M87 from the VLBA data at 24 GHz observed on 2018 Apr 28 (Project code: BG250A). The polarization structure near the core was similar to those at 43 GHz presented in that paper. We repeated the same analysis explained in this section and the following sections for the 24 GHz data, and the results are shown in Appendix~\ref{appendix:kband}. We could reproduce a similar core polarization image to that of the previous study when using LPCAL on 3C 279. However, we found several patches of weak polarization near the core in the GPCAL-processed image. The weak polarization could be due to stronger depolarization at lower frequencies \citep[e.g.,][]{Sokoloff1998}, which has been observed in many AGN jets \citep[e.g.,][]{OSullivan2012, OSullivan2017, Kravchenko2017, Park2018, Pasetto2018}. Since the observed polarized intensity is low and the structure is not fully consistent with that at 43 GHz, we conclude that more observations and archival data analysis are needed for obtaining convincing results at this frequency. Nevertheless, our synthetic data test and closure trace analysis show that the complex core polarization structure obtained from LPCAL using 3C 279 at this frequency may also be significantly affected by residual D-terms, similar to the results at 43 GHz.

\section{Synthetic data analysis}
\label{sec:synthetic}

It is not straightforward to tell which polarization image better represents the true polarization structure of M87 because the true D-terms are not known. In this section, we test the different calibration strategies by using synthetic data sets.

\begin{figure*}[t!]
\centering
\includegraphics[width = 1.0\textwidth]{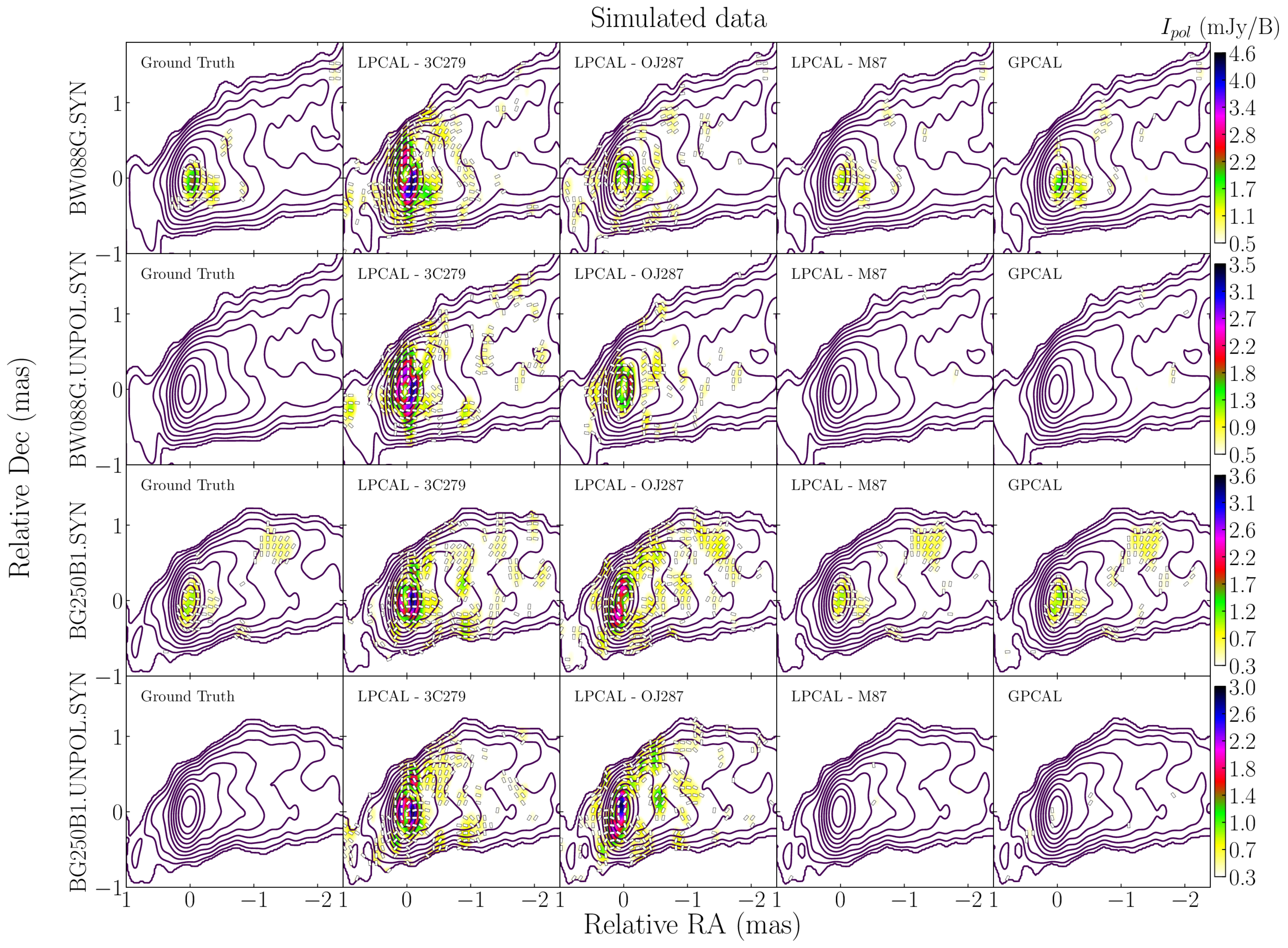}
\caption{Linear polarization images reconstructed from synthetic data sets. The ground-truth linear polarization images are presented in the leftmost column, and the reconstructed images from different calibration versions are presented in other columns. The upper and lower two rows show the results of synthetic data generated based on the BW088G and BG250B1 data, respectively. The second and fourth rows from the top show the results for the synthetic data set assuming no polarization for M87. \label{fig:syntheticdata}}
\end{figure*}

\begin{figure*}[t!]
\centering
\includegraphics[width = 1.0\textwidth]{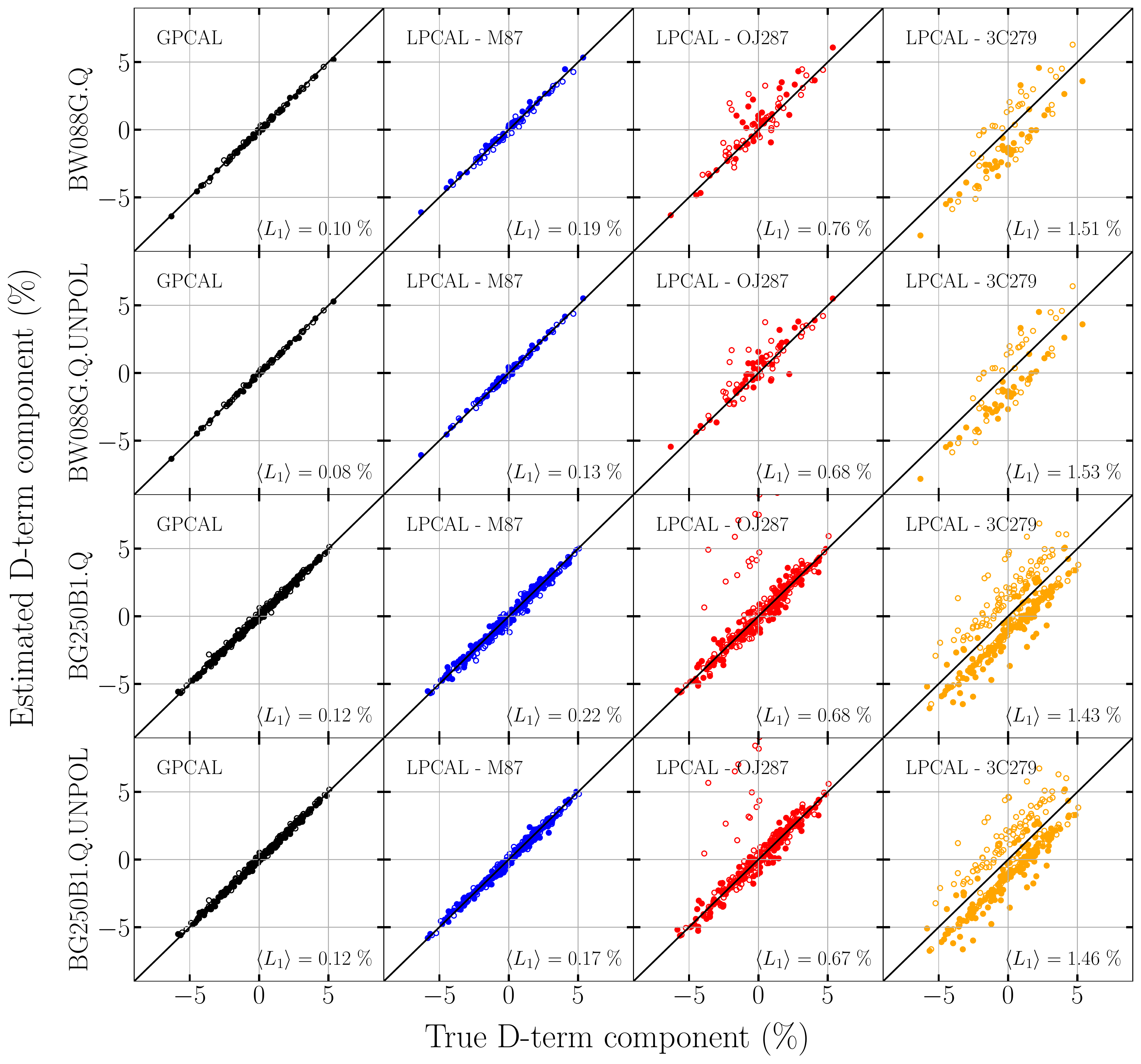}
\caption{Comparison of the ground-truth D-term components (real and imaginary parts) and reconstructed ones. The results for each synthetic data are presented in each row. Each column shows the results for each calibration version. The filled and open data points show the D-term components for RCP and LCP, respectively. The norm $L_1 \equiv |D-D_{\rm Truth}|$, averaged over left, right, real, and imaginary components of the D-terms and over all antennas and IFs, is noted at the bottom right. \label{fig:dtermcomp}}
\end{figure*}

We generated synthetic data with GPCAL. The equations used for generating the simulated visibilities for four cross correlation products on a baseline $mn$ ($r_{mn}^{RR}, r_{mn}^{RL}, r_{mn}^{LR}, r_{mn}^{LL}$) are
\begin{eqnarray}
\label{eq:meas}
&&r_{mn}^{RR} = G_m^{R}G_n^{R*}[\mathscr{RR} + D^R_me^{2j\phi_m} \mathscr{LR} + \nonumber\\ &&\quad D^{R*}_ne^{-2j\phi_n} \mathscr{RL} + D^R_{m}D^{R*}_{n}e^{2j(\phi_m-\phi_n)} \mathscr{LL}] \nonumber \\
&&r_{mn}^{RL} = G_m^{R}G_n^{L*}[\mathscr{RL} + D^R_me^{2j\phi_m} \mathscr{LL} + \nonumber\\&&\quad D^{L*}_ne^{2j\phi_n} \mathscr{RR} + D^R_{m}D^{L*}_{n}e^{2j(\phi_m+\phi_n)} \mathscr{LR}] \nonumber \\
&&r_{mn}^{LR} = G_m^{L}G_n^{R*}[\mathscr{LR} + D^L_me^{-2j\phi_m} \mathscr{RR} +  \nonumber\\&&\quad D^{R*}_ne^{-2j\phi_n} \mathscr{LL} + D^L_{m}D^{R*}_{n}e^{-2j(\phi_m+\phi_n)} \mathscr{RL}] \nonumber\\
&&r_{mn}^{LL} = G_m^{L}G_n^{L*}[\mathscr{LL} + D^L_me^{-2j\phi_m} \mathscr{RL} + \nonumber\\&&\quad D^{L*}_ne^{2j\phi_n} \mathscr{LR} + D^L_{m}D^{L*}_{n}e^{-2j(\phi_m-\phi_n)} \mathscr{RR}],
\end{eqnarray}
where the asterisk denotes a complex conjugate, $G$ is the complex antenna gain, $D$ is the leakage factor, and $\phi$ is the antenna field rotation angle. Subscripts denote antenna numbers, and superscripts denote polarization. The field rotation angle is equivalent to the parallactic angle for Cassegrain mounts, and thus for the VLBA antennas (see \citealt[][]{Park2021a} for more details). $\mathscr{RR}, \mathscr{RL}, \mathscr{LR}$, and $\mathscr{LL}$ are the true visibilities, which are related to the Fourier transforms of source's Stokes $I, Q, U$, and $V$ images ($\tilde{I}$, $\tilde{Q}$, $\tilde{U}$, and $\tilde{V}$) via the relations:
\begin{eqnarray}
\label{eq:stokes}
    \mathscr{RR} &=& \tilde{I} + \tilde{V}\nonumber\\
    \mathscr{RL} &=& \tilde{Q} + i\tilde{U}\nonumber\\
    \mathscr{LR} &=& \tilde{Q} - i\tilde{U}\nonumber\\
    \mathscr{LL} &=& \tilde{I} - \tilde{V}.
\end{eqnarray}
Equation~\ref{eq:meas} is identical to the standard Radio Interferometer Measurement Equations (RIMEs; \citealt{Hamaker1996, Smirnov2011}) except that it assumes that the field rotation angles were already corrected. The field rotation angles are usually corrected at an upstream calibration stage and both LPCAL \citep{Leppanen1995} and GPCAL \citep{Park2021a} use the measurement equations after the correction.

We assumed that the sources have no circular polarization signals and thus $\tilde{V}=0$. The model images used for the synthetic data generation are the Stokes $I$, $Q$, and $U$ CLEAN models obtained from the calibration using GPCAL (for all three sources; see the right panels of Figure~\ref{fig:realdata} for the M87 images). We added thermal noise to the synthetic data based on the observed visibility uncertainties. We assumed that the antenna gains are unity and the D-terms are constant during observations. The D-terms for corrupting the data were randomly chosen based on the D-term distribution of the real data estimated by GPCAL. We generated two synthetic data sets corresponding to two real data sets (BW088G and BG250B1), and each synthetic data set consists of three sources (M87, OJ 287, and 3C 279). The assumed D-terms are the same for all sources for each data set, but different between data sets.

We generated another version of synthetic data sets, assuming that M87 is unpolarized. This was to prevent us from being biased to the model from a particular calibration version (the GPCAL-processed images were assumed to be the true images for generating the synthetic data sets; see above). Any linear polarization structure reconstructed for these data sets should be attributed to imperfect instrumental polarization calibration.

We repeated the instrumental polarization calibration and polarimetric imaging procedures on the synthetic data sets as we did on the real data sets (Section~\ref{sec:vlba}). We present the ground truth and reconstructed polarization images for each calibration version in Figure~\ref{fig:syntheticdata}. The true images could not be reproduced when LPCAL is used on 3C 279 or OJ 287, while using LPCAL on M87 or using GPCAL could reproduce images similar to the true images. Interestingly, the images reconstructed by using LPCAL on 3C 279 or OJ 287 show significant linearly polarized intensity near the core even though M87 is assumed to be unpolarized (the second and fourth rows from the top in Figure~\ref{fig:syntheticdata}). The distributions of linearly polarized intensity and EVPAs are very similar to those from the real data sets obtained with the same calibration strategy (the second and fourth rows from the top in Figure~\ref{fig:realdata}). This result indicates that the complex linear polarization structure in the core of M87 shown in the previous studies obtained by using LPCAL on 3C 279 may be significantly affected by residual D-terms in the data.

In Figure~\ref{fig:dtermcomp}, we compare the ground truth D-terms with the reconstructed ones for each calibration version. We calculate the $L_1 \equiv |D_{i, \rm recon}-D_{i, \rm Truth}|$ norm, where $D_{i, \rm recon}$ is a reconstructed D-term component for a measurement $i$ and $D_{i, \rm Truth}$ the corresponding ground-truth D-term component. The $\langle L_1 \rangle$ value for each calibration version (averaged over antennas, IFs, D-term components) is noted in each panel. The calibrations using GPCAL and using LPCAL on M87 could reproduce the ground-truth D-terms at a level of $\langle L_1 \rangle \lesssim0.2\%$, while the reconstructed D-terms using LPCAL on OJ 287 and 3C 279 show large deviations from the true D-terms. $\langle L_1 \rangle$ is at a level of $\sim1.5\%$ when using LPCAL on 3C 279, which explains why the corresponding polarization images are quite different from the true images (Figure~\ref{fig:syntheticdata}). This result is not surprising. The historical and ongoing monitoring programs of many AGN jets using the VLBA at 43 GHz (the VLBA-BU-BLAZAR program and the BEAM-ME program\footnote{\url{https://www.bu.edu/blazars/VLBAproject.html}}) have found that the D-terms derived from LPCAL using 3C 279 significantly deviate from those using other good calibrators (Svetlana G. Jorstad, private communication).

The main reason why using LPCAL on 3C 279 or OJ 287 could not accurately reconstruct the true D-terms is due to a breakdown of the similarity approximation. Both sources show complex linear polarization structures, which are not proportional to their total intensity structures. In our previous study, we showed that even a very small positional shift ($\lesssim10\%$ of the synthesized beam size) between the total intensity and linear polarization peaks in the simulated data can prevent an accurate D-term reconstruction (Appendix B in \citealt{Park2021a}). We obtained a root-mean-square error of $\sim0.52\%$ in the reconstructed D-terms for the simulated data, which corresponds to $\langle L_1 \rangle \sim0.41\%$. However, the complex polarization structures are properly taken into account by iteratively solving for D-terms and improving the source polarization models with GPCAL, allowing us to obtain accurate D-term estimates and linear polarization images of M87.

\section{Model comparison using closure quantities}
\label{sec:ctrace}

We have demonstrated, using synthetic data sets, that instrumental polarization calibration using LPCAL on 3C 279 or OJ 287 can introduce artificial polarization structures in the M87 images due to residual D-terms. However, the synthetic data sets may not perfectly represent the real data sets as they were generated based on simple assumptions such as unity antenna gains, constant D-terms during observations, and no circular polarizations.

One can determine which reconstructed polarization image better represents the true source image by using closure traces, which are closure quantities constructed from parallel-hand and cross-hand visibilities that are insensitive to both antenna gain and D-term corruptions \citep{BP2020}. Subsets of these quantities are the well-known closure amplitudes and closure phases which do not depend on antenna gain corruption. We define the coherency matrix for a given pair of antennas, $A$ and $B$, as

\begin{equation}
\mathbf{V}_{AB} = 
\begin{pmatrix}
r^{RR}_{AB} & r^{RL}_{AB} \\
r^{LR}_{AB} & r^{LL}_{AB}
\end{pmatrix}.
\end{equation}

\noindent The closure trace on baselines connecting four antennas \{$A, B, C, D$\} can be obtained by

\begin{equation}
    \mathcal{T}_{ABCD} = \frac{1}{2}{\rm tr}(\mathbf{V}_{AB}\mathbf{V}^{-1}_{CB}\mathbf{V}_{CD}\mathbf{V}^{-1}_{AD}).
\end{equation}

\noindent This complex quantity is independent of antenna gains and D-terms. One can obtain another useful quantity called a conjugate closure trace product as

\begin{equation}
    \mathcal{C}_{ABCD} = \mathcal{T}_{ABCD}\mathcal{T}_{ADCB},
\end{equation}

\noindent which is identically unity in the absence of polarization. Thus, deviations from unity in the $\mathcal{C}_{ABCD}$ are a signature of source polarization that is independent on calibration. $\mathcal{C}_{ABCD}$ are much better constrained than either $\mathcal{T}_{ABCD}$ or $\mathcal{T}_{ADCB}$ as the constituent $\mathcal{T}$ are correlated \citep{BP2020}. Therefore, we use conjugate closure trace products for comparing models.

For a given quadrangle $ABCD$ (in the absence of autocorrelation quantities, see \citealt{BP2020}), there are six nonredundant complex $\mathcal{T}$: $\mathcal{T}_{ABCD}$, $\mathcal{T}_{ABDC}$, $\mathcal{T}_{ACBD}$, $\mathcal{T}_{ACDB}$, $\mathcal{T}_{ADBC}$, and $\mathcal{T}_{ADCB}$. One can obtain corresponding three nonredundant complex $\mathcal{C}$: $\mathcal{C}_{ABCD}$, $\mathcal{C}_{ABDC}$, and $\mathcal{C}_{ACBD}$. Thus, the number of "maximal set" of $\mathcal{C}$ for N antennas would be $3\binom{N}{4}$, while the majority of them are redundant with one another for large $N$. Unfortunately, selecting a minimal (non-redundant) set of closure traces for large $N$ is not straightforward\footnote{The methods for selecting minimal sets of closure phase and closure amplitude are presented in \cite{Thompson2017, Blackburn2020}}, and we will use the maximal set. Nevertheless, it should be enough to tell which model can better reproduce the observed $\mathcal{C}$.

We quantify agreement between a trial image and a (maximal) set of measured $\mathcal{C}$ using the mean squared standardized residual as

\begin{equation}
    \chi^2_{\mathcal{C}} = \frac{1}{2N_{\mathcal{C}}}\sum \frac{|\mathcal{C} - \hat{\mathcal{C}}|^2}{\sigma_{\mathcal{C}}^2},
\end{equation}

\noindent where $\hat{\mathcal{C}}$ denotes a model conjugate closure trace product obtained from the CLEAN models of the trial image, and the sum ranges over all measured $\mathcal{C}$ with the number of $N_{\mathcal{C}}$ (including three $\mathcal{C}$ for a given quadrangle). We used the data averaged over frequency (over all IFs) to enhance the S/Ns for the calculation. $\sigma_{\mathcal{C}}$ is the uncertainty estimated numerically through Monte Carlo sampling of the constituent visibilities \citep{BP2020}. The quantity $\chi^2_{\mathcal{C}}$ does not formally correspond to a reduced $\chi^2$ as we did not include a correction for the effective number of image degrees of freedom, and the observed ${\mathcal{C}}$ are not fully independent with each other (see, e.g., \citealt{EHT2019d}). We assumed that our sources have zero circular polarization, i.e., $\hat{V}=0$ and $\hat{I} = \mathscr{RR} = \mathscr{LL}$, for calculating $\hat{\mathcal{C}}$.

Ideally, one should compare $\chi^2_{\mathcal{C}}$ values for the M87 images to determine which polarization image is more reliable. However, M87 is very weakly polarized, and we could not find any noticeable deviation of the $\mathcal{C}$ values from unity. The resulting $\chi^2_{\mathcal{C}}$ values for different images are nearly identical to each other. Nevertheless, the calibrators, 3C 279 and OJ 287, have moderate levels of linear polarization \citep[e.g.,][]{Jorstad2017} and they are more suitable for determining which instrumental polarization calibration version achieves a better accuracy. If the calibrator polarization images obtained with GPCAL provide lower $\chi^2_{\mathcal{C}}$ values than those from LPCAL, it indicates that the D-terms from GPCAL are more accurate than those from LPCAL. Thus, the M87 polarization images obtained with GPCAL should better represents the true images than those with LPCAL. We selected the BG250B1 data for model comparison because of its higher S/N and greater discriminating power than the earlier epoch data sets\footnote{We also checked the BW088G data and found similar conclusions that the images obtained with GPCAL give lower $\chi^2_{\mathcal{C}}$ than those with LPCAL. However, the difference was not as significant as for the BG250B1 data, presumably because the BW088G data have a lower S/N due to a smaller bandwidth.}. We used the same Stokes I image when calculating $\chi^2_{\mathcal{C}}$ to make sure that the difference in $\chi^2_{\mathcal{C}}$ between models can solely be determined by the linear polarization images of the models and the D-term calibration accuracy.

\begin{figure}[t!]
\centering
\includegraphics[width = 0.47\textwidth]{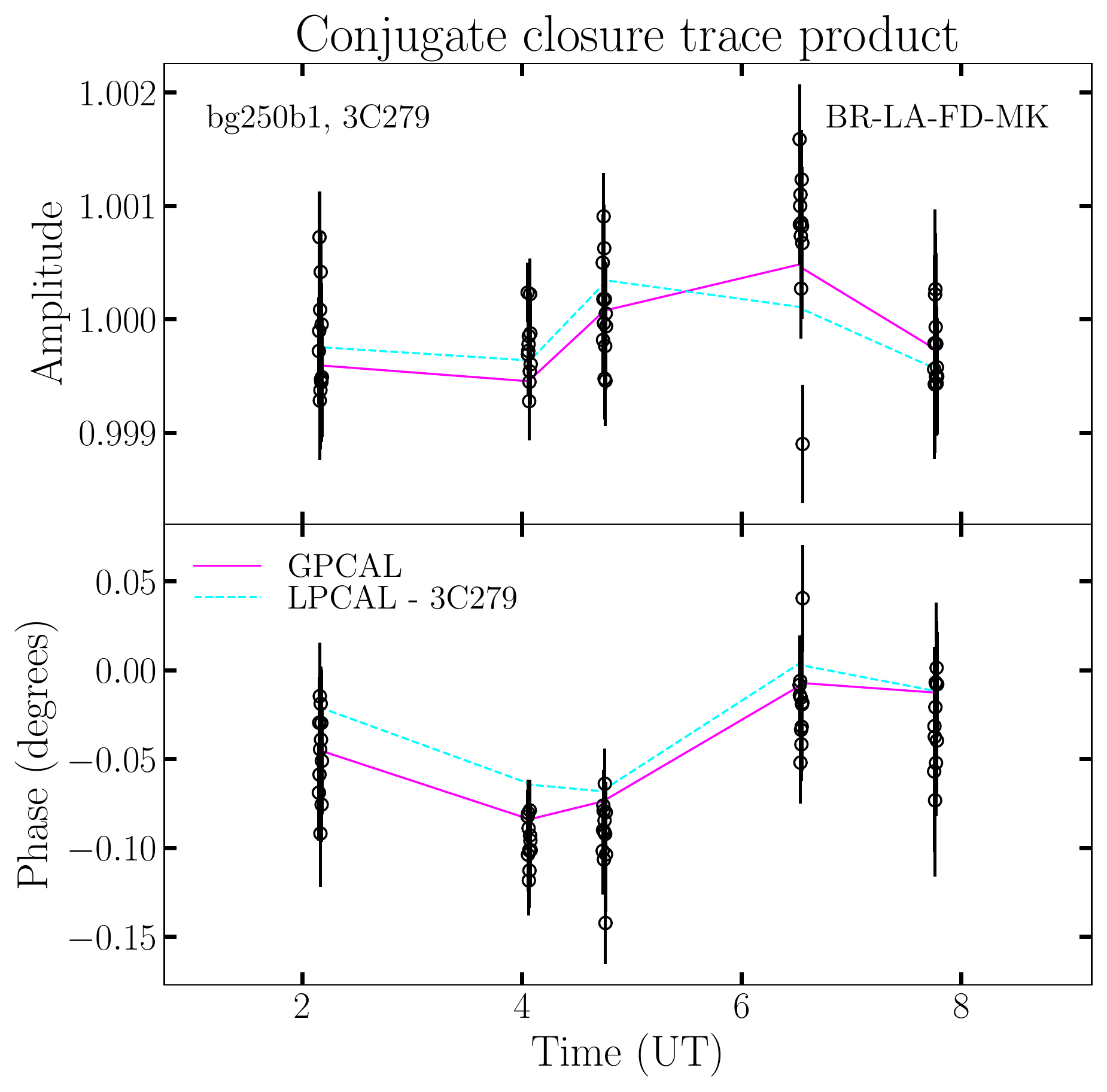}
\caption{Amplitude and phase of conjugate closure trace products constructed on the BR-LA-FD-MK quadrangle for 3C 279 in the BG250B1 data. The magenta and cyan dashed lines show the same conjugate closure trace products for the 3C 279 images reconstructed by using GPCAL and LPCAL on 3C 279, respectively. \label{fig:ctrace}}
\end{figure}

\begin{deluxetable}{ccccc}
\tablecaption{$\chi^2_{\mathcal{C}}$ for images of 3C 279 and OJ 287 \label{tab:ctrace}}
\tablewidth{0pt}
\tablehead{ & \colhead{GPCAL} & \colhead{LPCAL$_{\rm M87}$} & \colhead{LPCAL$_{\rm OJ287}$} & \colhead{LPCAL$_{\rm 3C279}$}}
\startdata
$\chi^2_{C, {\rm 3C279}}$ & 0.90 & 0.98 & 1.67 & 0.95 \\
$\chi^2_{C, {\rm OJ287}}$ & 0.91 & 0.91 & 0.93 & 0.95 \\
\enddata
\tablecomments{Values in each column show $\chi^2_{\mathcal{C}}$ for images for each calibration strategy. $\chi^2_{\mathcal{C}}$ is obtained by using the BG250B1 data after averaging over frequency.}
\end{deluxetable}

\begin{figure*}[t!]
\centering
\includegraphics[trim = 3mm 4mm 2mm 1mm, width = 1.0\textwidth]{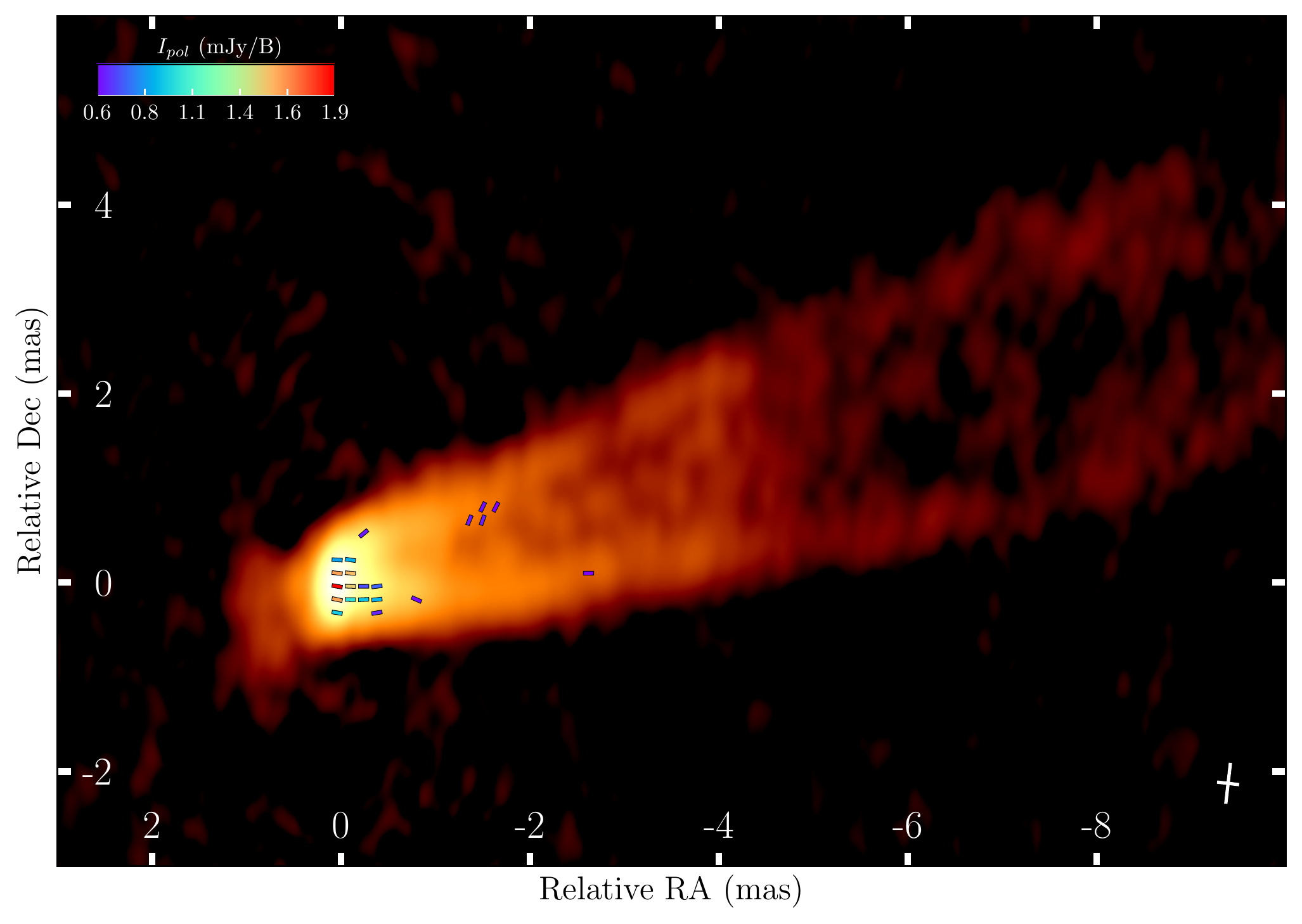}
\caption{4-epoch average polarimetric image of the M87 jet. The image is the average of the GPCAL-processed images convolved with the average synthesized beam with the shape of $0.407\times0.204{\rm\ mas}, -6.53^\circ$ (shown with the white crosses in the bottom right corner). The colored ticks show EVPAs. \label{fig:stack}}
\end{figure*}

In Table~\ref{tab:ctrace}, we present $\chi^2_{\mathcal{C}}$ for source images obtained with different calibration strategies. The $\chi^2_{\mathcal{C}}$ values for the images obtained with GPCAL are smaller than those with LPCAL using 3C 279 and OJ 287 for both sources. In Figure~\ref{fig:ctrace}, we present the observed and model conjugate closure trace products for the BR-LA-FD-MK quadrangle as an example. The GPCAL model (magenta) describes the observed data better than the LPCAL-3C279 model (cyan) as the statistics suggests, especially the phases. Nevertheless, the latter model still fits to the data reasonably well. This implies that the D-term calibration using LPCAL on 3C 279 is still reasonably good but it achieves a less accuracy than GPCAL. However, this slightly poor accuracy can significantly affect the linear polarization images of M87 because it has a very low fractional polarization of $\sim0.2$--$0.6\%$ at the core. 

We found that the LPCAL estimates from OJ 287 are more consistent with the GPCAL estimates as compared with the LPCAL estimates from 3C 279, except for the D-terms of SC antenna. The large deviation for the SC D-terms is due to a smaller parallactic angle coverage of OJ 287 ($\approx20^\circ$) than 3C 279 ($\approx50^\circ$). The impact of the inaccurate SC D-terms on the source polarization models can be more significant for 3C 279 than OJ 287, because the latter is more core-dominated and SC antenna comprises long baselines. This could be the reason why we obtained the particularly high $\chi^2_{\mathcal{C}, {\rm 3C 279}}$ value when using LPCAL on OJ 287 (see also Appendix~\ref{appendix:dtermstability}).

The result of our conjugate closure trace product analysis indicates that GPCAL obtains more accurate D-terms than LPCAL using 3C 279 or OJ 287. Therefore, we conclude that the GPCAL polarization images of M87 showing a compact core polarization structure is closer to the true image than the LPCAL images obtained in previous studies \citep{Walker2018, Kravchenko2020}.

\section{Discussion}
\label{sec:discussion}

We found that the linear polarization in the core of the M87 jet at 43 GHz is characterized as a compact structure having fractional polarizations of $\sim0.2-0.6\%$ and EVPAs of $\sim66-92$ degrees depending on epoch (Table~\ref{tab:data}). These quantities show a moderate level of variability, similar to the variability of the core in total intensity observed in the stable periods (\citealt{Walker2018}; see \citealt{Acciari2009, Hada2014} for flaring at the core in this source). The mean and standard deviation of the observed EVPAs are $79.4\pm12.4^\circ$.

In Figure~\ref{fig:stack}, we present the average polarimetric image of the GPCAL-processed images in four epochs. The Stokes $I$, $Q$, and $U$ models of individual epoch data are convolved with the average synthesized beam with the shape of $0.407\times0.204{\rm\ mas}, -6.53^\circ$ before averaging. The total intensity jet and counterjet structures are very similar to the 23-epoch average image presented in \cite{Walker2018}. The linear polarization of the jet is the most bright at the core. An elongated polarization structure along the southern jet limb connected to the core polarization is also observed. In the outer jet region, a few patches with weak polarization are seen. Interestingly, the observed EVPAs near the core seem to be well aligned with the global jet direction. This behavior is reminiscent of the good alignment between the core EVPAs and the inner jet direction in BL Lacertae objects \citep[e.g.,][]{LH2005, Jorstad2007, Hodge2018}.

\begin{figure}[t!]
\centering
\includegraphics[width = 0.47\textwidth]{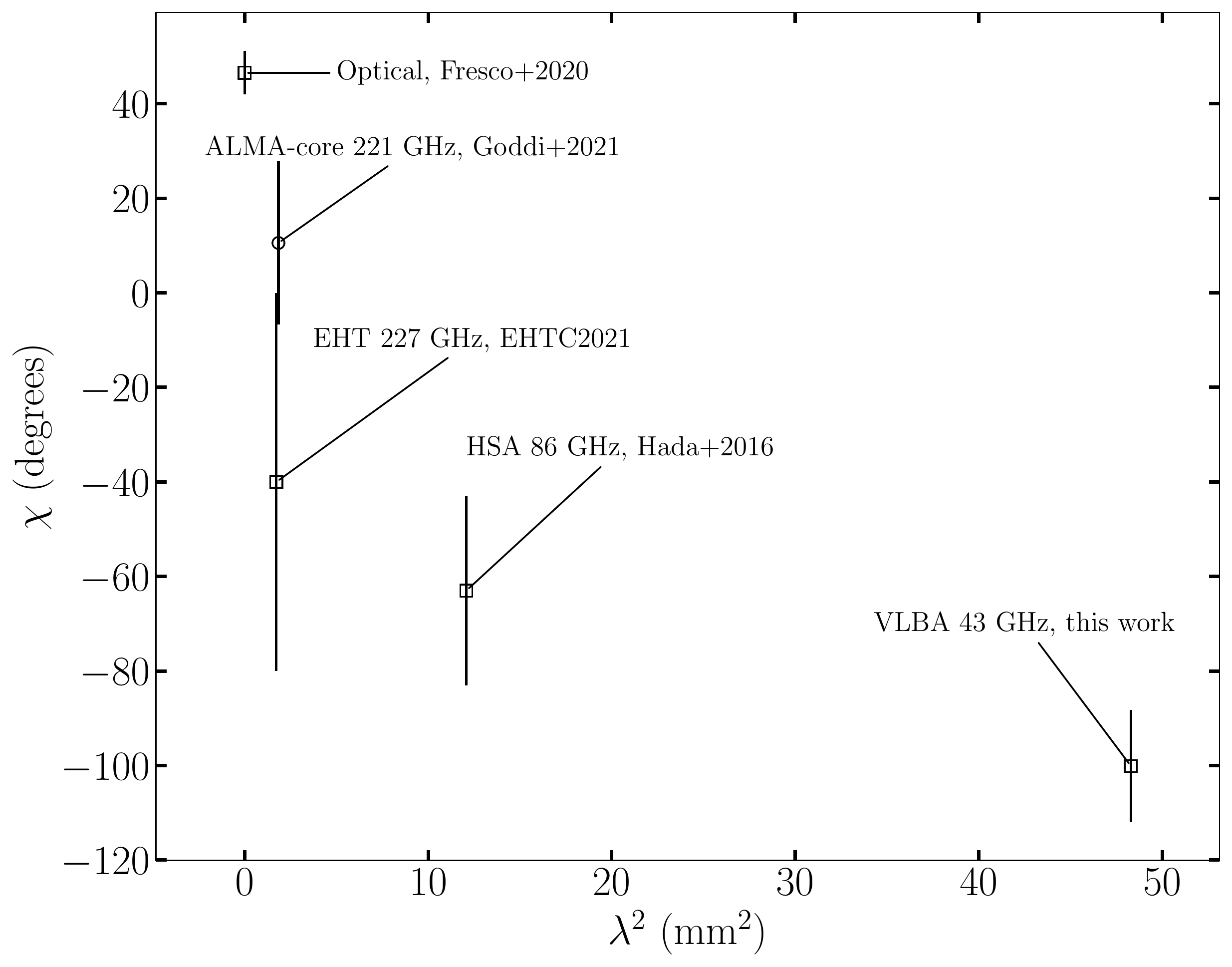}
\caption{EVPA as a function of $\lambda^2$. The VLBA 43 GHz data point is the average of the core EVPAs in four epochs. The HSA 86 GHz data point refers to the linear polarization blob at $\sim0.1$ mas downstream the core presented in \cite{Hada2016}. The EHT 227 GHz data point and error bar show the representative value and the uncertainty range of the net EVPA observed on horizon scales of M87 in April 2017 \citep{EHT2021a}. The ALMA-core 221 GHz data point presents the mean and standard deviations of the mean-wavelength EVPAs at the nucleus of M87 obtained from observations of M87 with the Atacama Large Millimeter/submillimeter Array (ALMA) in 2017 and 2018 \citep{Goddi2021}. The optical data point is the weighted sum of the $I$ and $V$-band EVPAs measured at the nucleus of M87 observed with the HST \citep{Fresco2020}. \label{fig:evpas}}
\end{figure}

One must take into account Faraday rotation of observed EVPA to infer the intrinsic magnetic field geometry. It has been challenging to detect the linear polarization in the subparsec core at other frequencies, which is not surprising given the very low fractional polarization. The core is unpolarized at frequencies $\lesssim15$ GHz \citep{ZT2002, Park2019a}. \cite{Hada2016} reported $\chi\sim117\pm20^\circ$ at $\sim0.1$ mas downstream of the core at 86 GHz based on a high sensitivity array (HSA) observation. The recent observations of M87 with the EHT at 227 GHz have shown complex linear polarization structures on event horizon scales \citep{EHT2021a}. A conservative range for the net EVPA integrated over the images can be described as $\chi=-80$--$0^\circ$, which depends on analysis method, frequency, and observation date. More measurements at 1 mm are available from the ALMA-only data \citep{Goddi2021}. We take the mean and standard deviations of the reported, mean-wavelength EVPAs (averaged over the frequency range of 212--230 GHz; see Table 3 in \citealt{Goddi2021}) in the nucleus of M87 observed in 2017 and 2018, which is $\chi\sim10.5\pm17.2^\circ$. At optical wavelengths, \cite{Fresco2020} detected linear polarization at the nucleus of the M87 jet with the Hubble Space Telescope (HST), providing $\chi = 46.5\pm4.6^\circ$ (obtained by taking a weighted average of the $I$ and $V$ band measurements).

We present the core EVPAs obtained in our study and in the literature as a function of $\lambda^2$ in Figure~\ref{fig:evpas}. Unfortunately, it is impossible to constrain the RM and intrinsic EVPA in the core with these measurements only. Firstly, the ALMA and HST cannot resolve the subparsec-scale core, and the observed polarization can be contaminated by the extended jet polarization emission. Secondly, the EHT 227 GHz polarization is observed on horizon scales, where the emission at 43 or 86 GHz is expected to be optically thick \citep[e.g.,][]{Nakamura2018, Chael2019, EHT2019e}. Thus, the polarized emission at different frequencies observed with the VLBI arrays may pass through different parts of the Faraday screen, and one cannot constrain the RM with these measurements. Thirdly, it is unclear whether the 86 GHz polarization structure located at $\sim0.1$ mas downstream of the core and the 43 GHz core polarization structure originate in the same emitting region. The apparent difference in the polarization locations may be attributed to either the limited angular resolution of the VLBA at 43 GHz or possible contribution from non-negligible residual D-terms in the HSA data at 86 GHz. Lastly, we ignored time variability even though our data suggest moderate variability in polarization at 43 GHz on monthly time scales. The measurements in Figure~\ref{fig:evpas} span more than 10 years in time.

We plan to perform quasi-simultaneous multifrequency polarization observations of M87 in the near future to reveal the Faraday rotation and the intrinsic magnetic field geometry in the core. This will complement the ongoing and planned EHT observations which will cover a wide frequency range between 212 and 230 GHz \citep{EHT2021a}. This joint effort will reveal the systematic evolution of RM and magnetic field structure of the jet from horizon scales to downstream jet regions for the first time, enabling to test the models of jet launching and acceleration \citep[e.g.,][]{VK2004, McKinney2006, Tchekhovskoy2011}.

\section{Conclusions}
\label{sec:conclusion}

We have presented new linear polarization images of M87 based on VLBA data sets at 43 GHz observed in four epochs. These data sets were analyzed in previous studies \citep{Walker2018, Kravchenko2020}, which showed complex linear polarization structures surrounding the subparsec-scale core. We were motivated by two points and revisited the analysis. One is how the linear polarization structure can be such complex near the core, where total intensity emission is expected to be optically thick and very compact \citep[e.g.,][]{Hada2011, Hada2013}. The other is that the observed fractional polarization was quite low. In this case, the core polarization can be significantly affected by residual instrumental polarization signals in data \citep{Leppanen1995}.

We used GPCAL, a recently developed instrumental polarization calibration pipeline \citep{Park2021a}, for data calibration. There are two main advantages of GPCAL over the conventional tool LPCAL in AIPS, used for data calibration in the previous M87 studies \citep{Walker2018, Kravchenko2020}. One is that it allows to take into account resolved linear polarization structures of calibrators for D-term estimation. LPCAL assumes that the calibrators' linear polarization and total intensity structures are similar \citep{Cotton1993, Leppanen1995}, which does not always hold especially at high-frequencies. The other advantage is that GPCAL can fit the radio interferometric measurement equations to data from multiple calibrator sources simultaneously to enhance the calibration accuracy.

We found that our new linear polarization images of M87 show a compact structure in the core with its peak nearly coincident with the total intensity peak in all four epochs. The observed fractional polarizations and EVPAs at the core are $\approx0.35\%$ and $\approx79^\circ$, respectively, with an indication of mild variability over epochs. In addition, we obtained linear polarization images with calibration using LPCAL on individual calibrators (3C 279, OJ 287, and M87). The images obtained from LPCAL using 3C 279 are similar to those presented in the previous studies \citep{Walker2018, Kravchenko2020}. We claim that this difference originates from inaccurate instrumental polarization calibration using LPCAL, which can be attributed to the complex linear polarization structures of the calibrator that violate the similarity approximation in LPCAL.

We tested our claim by using synthetic data sets generated with GPCAL based on the real data sets observed in two epochs. We assumed that there is no antenna gain corruption and D-terms of each antenna are constant during observations. We used the linear polarization images of M87 and the calibrators, reconstructed by using GPCAL, as the ground-truth source polarization images of the synthetic data sets. We generated additional synthetic data sets by assuming that M87 is unpolarized. We found that the reconstructed linear polarization images of M87 using LPCAL on 3C 279 are similar to the images obtained from the real data sets with the same calibration strategy. This was the case even for the synthetic data sets assuming no polarization for M87. The reconstructed D-terms from LPCAL using 3C 279 and OJ 287 significantly deviate from the true values. Thus, we conclude that the complex polarization structures of M87 obtained by using LPCAL on 3C 279 and OJ 287 are significantly affected by residual D-terms. These calibrators show moderate to high levels of linear polarization and complex polarization structures, and it is not surprising that using LPCAL on those calibrators cannot achieve a calibration accuracy better than the assumed source polarization fraction for M87 ($\lesssim0.4\%$).

To verify if our new polarization images better represent the true polarization image of M87, we analyzed closure traces, which are quantities insensitive to antenna gain and polarimetric leakage corruptions \citep{BP2020}. We compared conjugate closure trace products of the calibrators, which are sensitive only to structures in the source polarization fraction, constructed from the data and the source images (models). We found that the images obtained from GPCAL better explain the observed conjugate closure trace products of the calibrators than the LPCAL images. This result suggests that the D-terms from GPCAL are more accurate than those from LPCAL, so are the M87 polarization images.

We compared the average core EVPA at 43 GHz with previous observations at higher frequencies. Unfortunately, it is prevented to constrain the RM and intrinsic EVPA in the core with these measurements only because they probe different spatial scales and time. We discuss the prospect for future multifrequency VLBI observations, which will constrain the evolution of the Faraday rotation and magnetic field structure from the jet base to the extended jet regions for the first time.

Most AGN jets that have been extensively investigated in polarization so far are of blazars. They usually show moderate to high levels of polarization \citep[e.g.,][]{Jorstad2007, Casadio2017, Hodge2018, Park2018, Park2019c}. Also, VLBI polarization images of blazars from large survey observations \citep[e.g.,][]{Jorstad2017, Lister2018} benefit from a number of sources in the same observing run. Those programs compare the D-terms from LPCAL using individual sources, discard outliers, obtain the average D-terms, and achieve a good calibration accuracy. Thus, the situation dealt with in this paper, having a low fractional source polarization $\lesssim0.5\%$ at the core and a limited number of sources in data, has not been encountered much previously. With GPCAL, however, one can achieve the D-term estimation accuracy good enough to detect the weak source polarization even though there are only a few (or even a single) calibrators in the data (see also Appendix~\ref{appendix:bm413i}). Our study suggests that polarization imaging of very weakly polarized sources has become possible with the advanced instrumental polarization calibration technique provided by GPCAL. 

\acknowledgments

We thank the anonymous ApJ referee for detailed comments that improved the manuscript. J.P. thanks Dominic W. Pesce, Svetlana G. Jorstad, and R. Craig Walker for fruitful discussions on closure traces and instrumental polarization calibration of VLBI data. J.P. acknowledges financial support through the EACOA Fellowship awarded by the East Asia Core Observatories Association, which consists of the Academia Sinica Institute of Astronomy and Astrophysics, the National Astronomical Observatory of Japan, Center for Astronomical Mega-Science, Chinese Academy of Sciences, and the Korea Astronomy and Space Science Institute. This work is supported by the Ministry of Science and Technology of Taiwan grant MOST 109-2112-M-001-025 and 108-2112-M-001-051 (K.A). M.K. acknowledges financial support from JSPS KAKENHI grant JP18K03656, JP18H03721, and JP21H01137. H.-Y. P. acknowledges the support of the Ministry of Education (MoE) Yushan Young Scholar Program, the Ministry of Science and Technology (MOST) under the grant 110-2112-M-003-007-MY2, and National Taiwan Normal University. E.V.K. is supported by the Russian Science Foundation (project 20-72-10078). The VLBA is an instrument of the National Radio Astronomy Observatory. The National Radio Astronomy Observatory is a facility of the National Science Foundation operated by Associated Universities, Inc. We are grateful to the staff of the KVN who helped to operate the array and to correlate the data. The KVN and a high-performance computing cluster are facilities operated by the KASI (Korea Astronomy and Space Science Institute). The KVN observations and correlations are supported through the high-speed network connections among the KVN sites provided by the KREONET (Korea Research Environment Open NETwork), which is managed and operated by the KISTI (Korea Institute of Science and Technology Information).

\facilities{VLBA (NRAO), VLA (NRAO), KVN (KASI)}

\software{AIPS \citep{Greisen2003}, Difmap \citep{Shepherd1997}, GPCAL \citep{Park2021a}}

\appendix

\section{Linear polarization images of M87 obtained with GPCAL without using M87}
\label{appendix:noM87}

\begin{figure*}[t!]
\centering
\includegraphics[width = 1.0\textwidth]{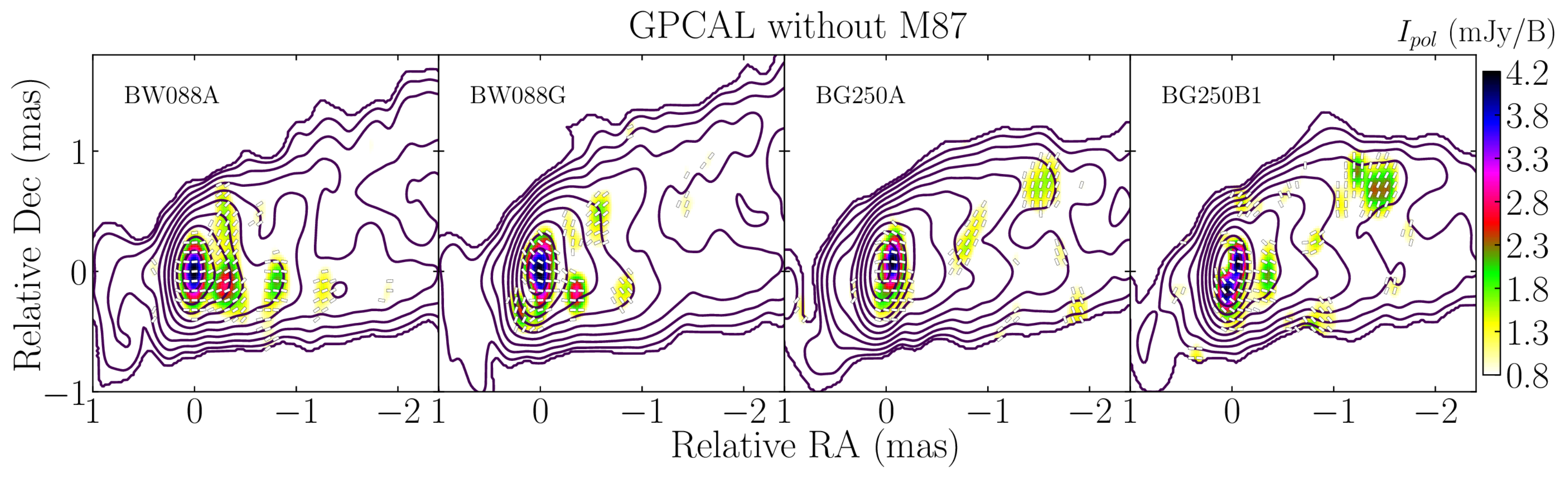}
\caption{Linear polarization images of M87 obtained with GPCAL using 3C 279 and OJ 287 only. \label{fig:noM87}}
\end{figure*}

In Section~\ref{sec:vlba}, the GPCAL pipeline used M87 for the initial D-term estimation assuming that it is unpolarized and 3C 279 and OJ 287 for instrumental polarization self-calibration. One may wonder if the initial D-term estimation from M87 can affect the following solutions significantly. We ran the pipeline again using 3C 279 and OJ 287 for the initial D-term estimation as well. The resulting linear polarization images of M87 are presented in Figure~\ref{fig:noM87}, which are very similar to the GPCAL images shown in Figure~\ref{fig:realdata}.

\section{Results of 24 GHz data analysis}
\label{appendix:kband}

In this appendix, we present the results of our 24 GHz data analysis. \cite{Kravchenko2020} presented the linear polarization image of M87 at 24 GHz based on the VLBA data (project code of BG250A) observed on 2018 Apr 28. The core polarization structure was similar to those at 43 GHz presented in that study. 

\begin{figure*}[t!]
\centering
\includegraphics[width = 1.0\textwidth]{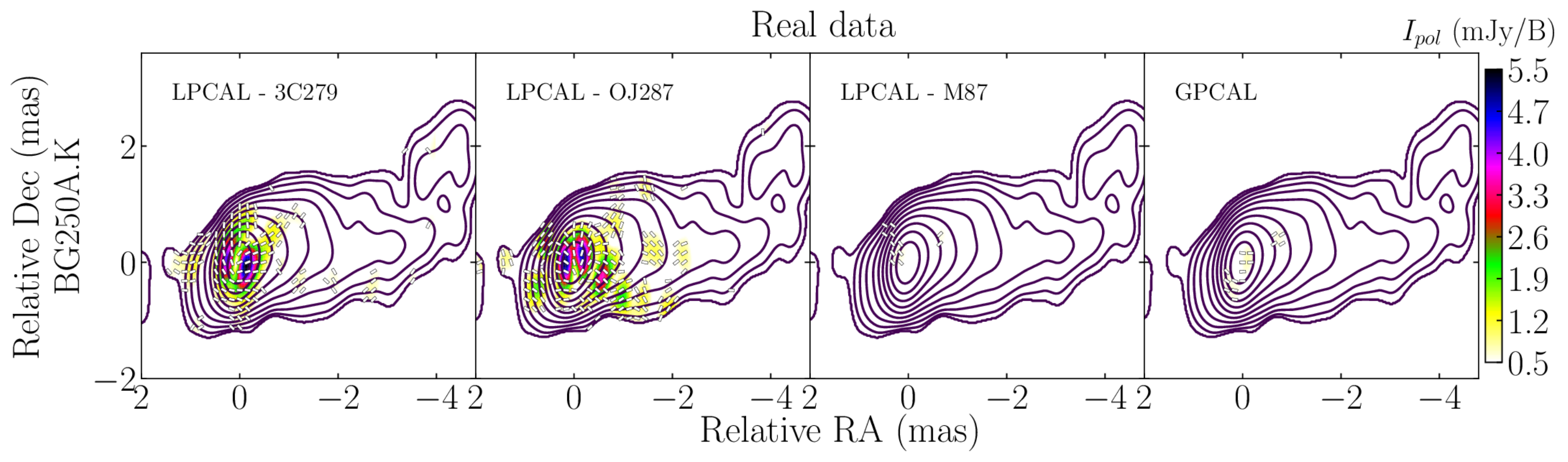}
\caption{Same as Figure~\ref{fig:realdata} but for the VLBA 24 GHz data. \label{fig:apprealdata}}
\end{figure*}

We have performed all the analysis we did in this paper for the 24 GHz data. Figure~\ref{fig:apprealdata} shows the linear polarization images of M87 reconstructed with different calibration strategies. The image obtained with LPCAL using 3C 279 presents two blobs having nearly perpendicular EVPA orientations to each other, which is consistent with the result of the previous study. The image obtained with LPCAL using OJ 287 also shows a complex polarization structure. However, we detected a polarization patch having much weaker intensity near the core in the image obtained with LPCAL using M87. A few weak polarization patches near the core are seen in the GPCAL-processed image.

\begin{figure*}[t!]
\centering
\includegraphics[width = 1.0\textwidth]{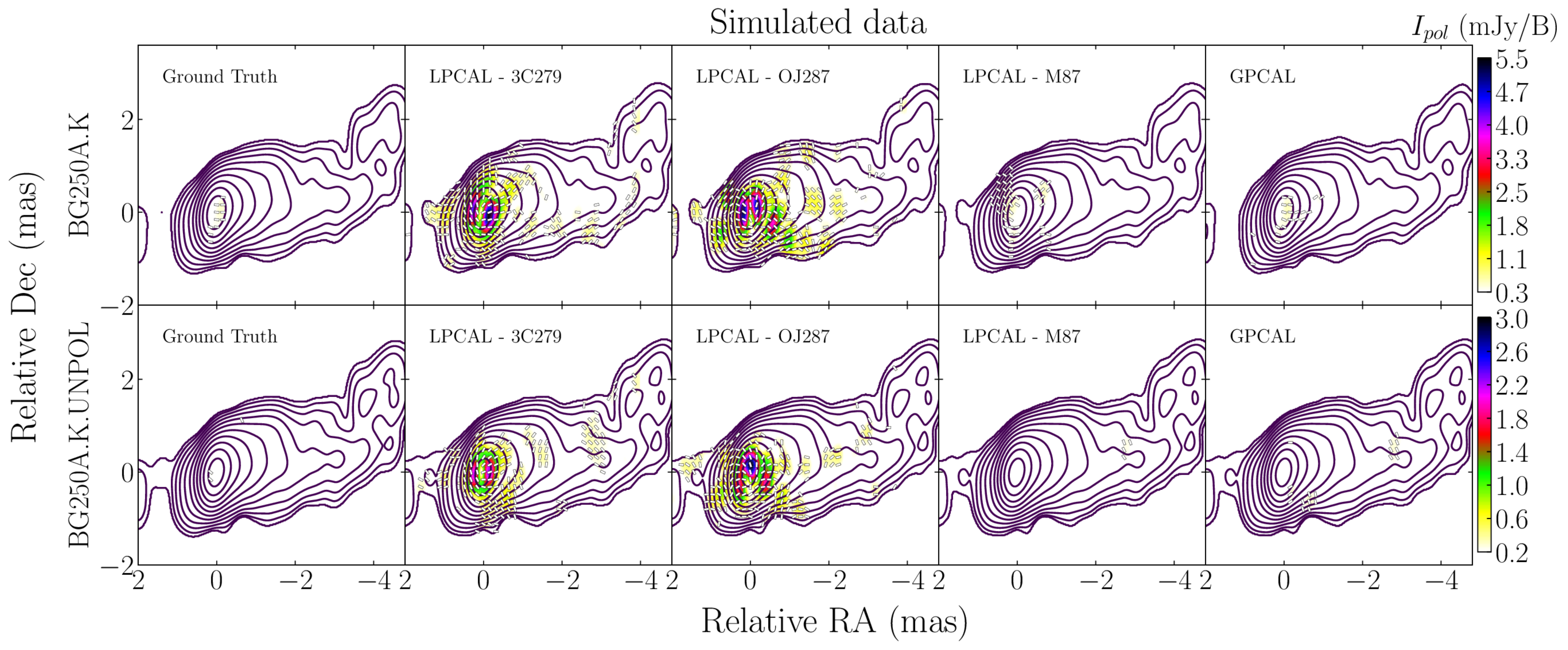}
\caption{Same as Figure~\ref{fig:syntheticdata} but for the VLBA 24 GHz data. \label{fig:appsyntheticdata}}
\end{figure*}

\begin{figure*}[t!]
\centering
\includegraphics[width = 1.0\textwidth]{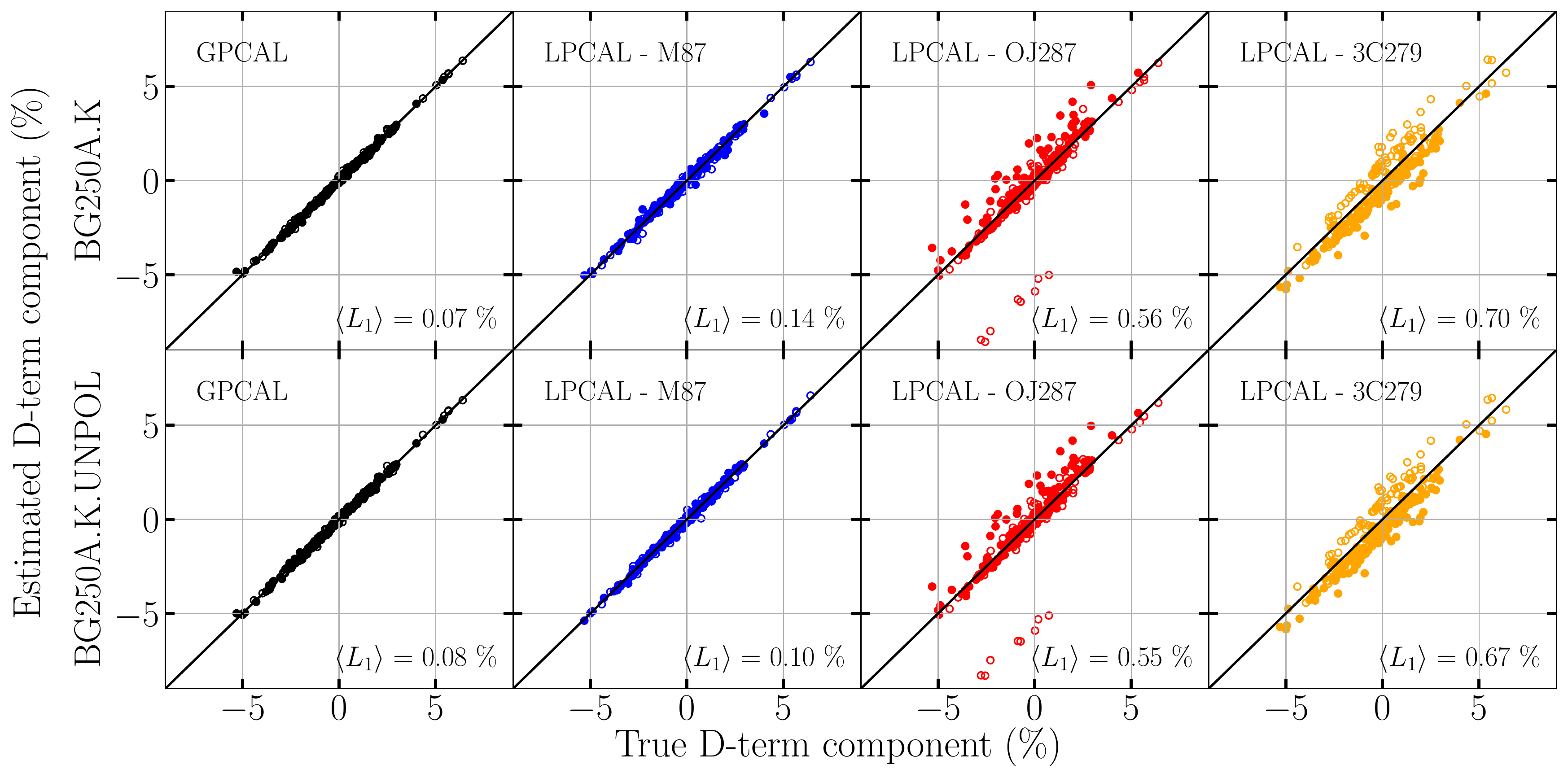}
\caption{Same as Figure~\ref{fig:dtermcomp} but for the VLBA 24 GHz data. \label{fig:appdtermcomp}}
\end{figure*}

In Figures~\ref{fig:appsyntheticdata} and ~\ref{fig:appdtermcomp}, we present the results of our synthetic data test for the 24 GHz data. We obtained similar results to those at 43 GHz. The reconstructed images from LPCAL using 3C 279 and OJ 287 are considerably different from the true images. Notably, the image reconstructed with LPCAL using 3C 279 from the synthetic data assuming no M87 polarization is consistent with the image from the real data obtained with the same calibration strategy. The reconstructed D-terms from LPCAL using 3C 279 and OJ 287 substantially deviate from the true D-terms, while those from LPCAL using M87 and from GPCAL are in good agreement with the true ones. Therefore, we conclude that the polarization image at 24 GHz presented in \cite{Kravchenko2020} could also be significantly affected by residual D-terms, which results from a breakdown of the similarity approximation in LPCAL.

In Table~\ref{tab:appctrace}, we present $\chi^2_{\mathcal{C}}$ for images of the calibrators from different calibration versions. Similar to our findings for the 43 GHz data, $\chi^2_{\mathcal{C}}$ for the GPCAL-processed images is much smaller than for the LPCAL-processed images for both calibrators. This result is consistent with our conclusion from the synthetic data test. However, the image obtained with GPCAL shows quite weak polarization, and the structure is not fully consistent with the 43 GHz results. The weak polarized intensity could be due to stronger depolarization at lower frequencies \citep[e.g.,][]{Sokoloff1998}, which has been observed in many AGN jets \citep[e.g.,][]{OSullivan2012, OSullivan2017, Kravchenko2017, Park2018, Pasetto2018}. We conclude that further investigation with more observations and archival data analysis is needed for obtaining convincing polarization structures near the core at this frequency.

\begin{deluxetable}{ccccc}
\tablecaption{$\chi^2_{\mathcal{C}}$ for images of 3C 279 and OJ 287 \label{tab:appctrace}}
\tablewidth{0pt}
\tablehead{ & \colhead{GPCAL} & \colhead{LPCAL$_{\rm M87}$} & \colhead{LPCAL$_{\rm OJ287}$} & \colhead{LPCAL$_{\rm 3C279}$}}
\startdata
$\chi^2_{C, {\rm 3C279}}$ & 1.75 & 2.50 & 5.71 & 2.62 \\
$\chi^2_{C, {\rm OJ287}}$ & 1.94 & 3.11 & 2.98 & 3.36 \\
\enddata
\tablecomments{Same as Table~\ref{tab:ctrace} but for the VLBA 24 GHz data.}
\end{deluxetable}

\section{D-term stability}
\label{appendix:dtermstability}

The D-terms of the VLBA are known to vary slowly, i.e., stable on monthly time scales\footnote{We note that sudden large changes in D-terms can occur when changes are made to the electronics or receivers at individual antennas. Thus, comparisons of the observations separated more than several months should be viewed cautiously.} \citep[e.g.,][]{Gomez2002}. Therefore, the stability of D-terms can be used as a test of the D-term estimation accuracy. If the LPCAL estimates are significantly affected by a breakdown of the similarity approximation as we claimed above, they are expected to be less stable over time than the GPCAL estimates. This is because the total intensity and linear polarization structures of both 3C 279 and OJ 287 are variable \citep[e.g.,][]{Jorstad2017}, which can affect the resulting LPCAL estimates. The GPCAL estimates should be much less affected by the variability.

\begin{figure*}[t!]
\centering
\includegraphics[width = 0.49\textwidth]{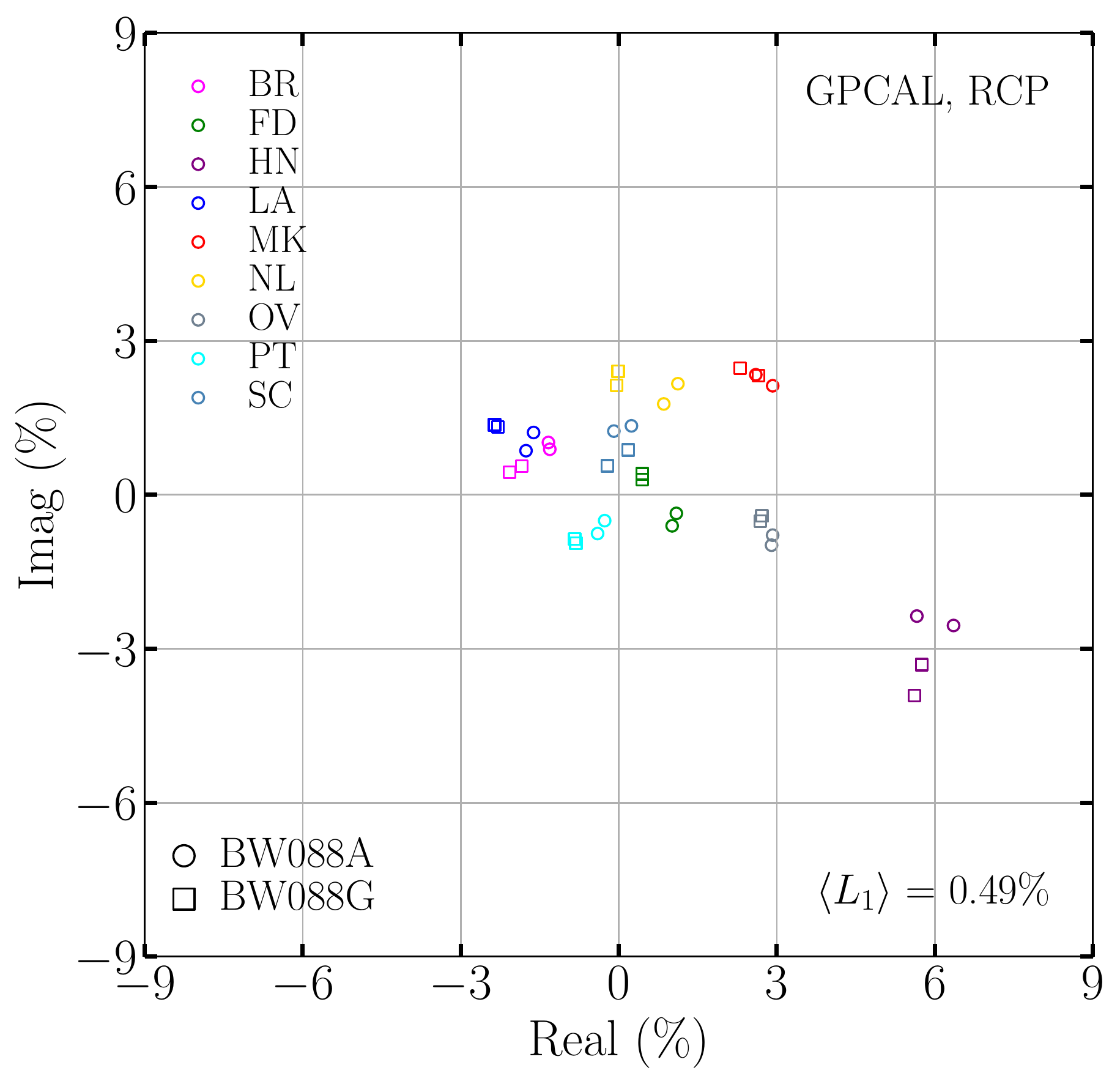}
\includegraphics[width = 0.49\textwidth]{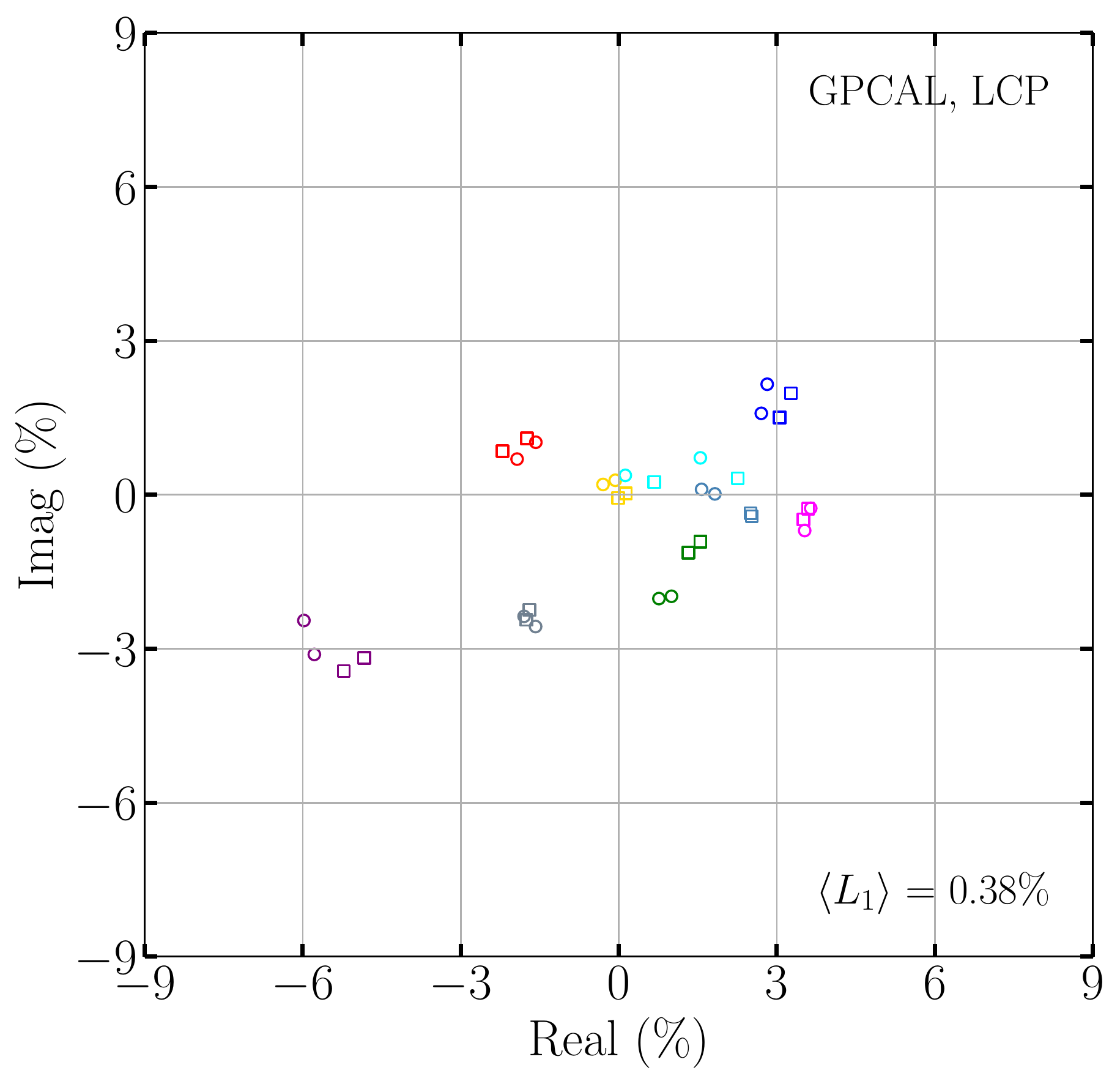}
\includegraphics[width = 0.49\textwidth]{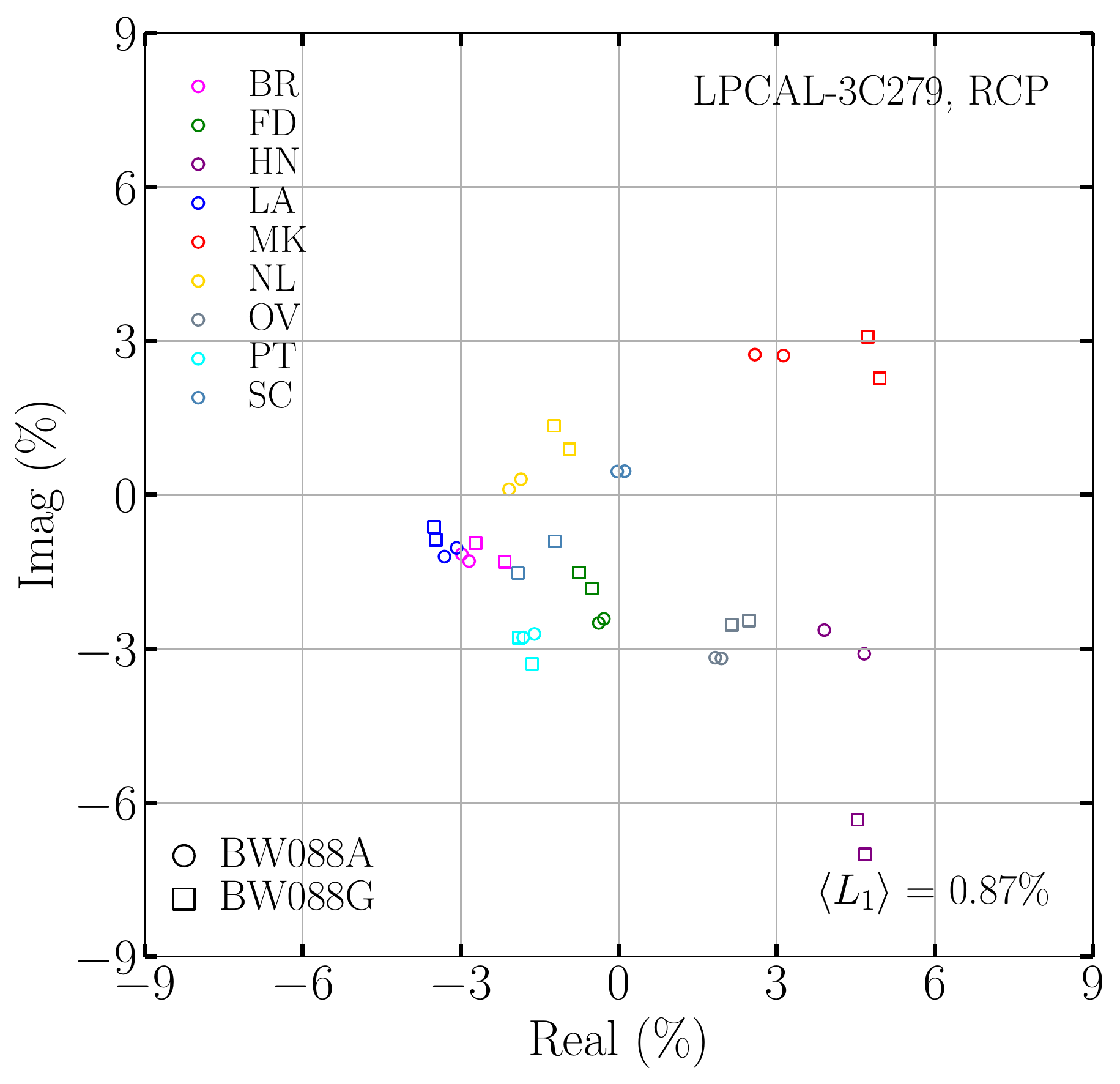}
\includegraphics[width = 0.49\textwidth]{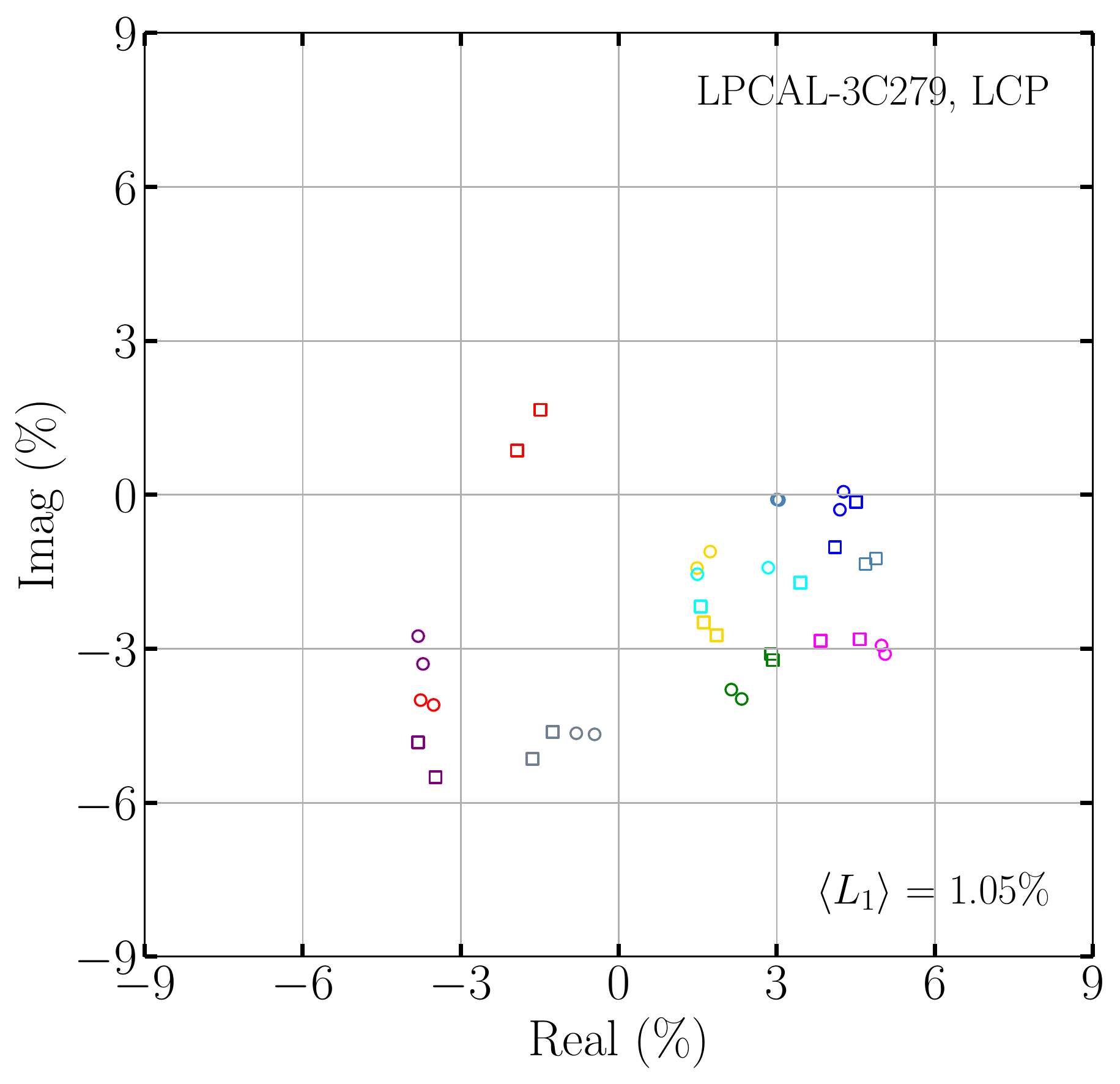}
\caption{Distribution of D-term estimates for RCP (left) and LCP (right) obtained with GPCAL (upper) and LPCAL using 3C 279 (lower). The circles and squares show the results for the BW088A and BW088G data, respectively. The average $L_1$ norm of the D-term estimates between epochs is noted in the bottom right corner of each panel. \label{fig:dtermstability}}
\end{figure*}

In Figure~\ref{fig:dtermstability}, we present the D-terms after subtracting the phase terms associated with the phase offset between polarizations at the reference antenna (see, e.g., footnote 9 in \citealt{Park2021a}). We found that the D-terms obtained by using GPCAL are quite stable with the average $L_1$ norm of the estimates between epochs of $\approx0.49, 0.38\%$ for RCP and LCP, respectively. However, we found larger deviations in the D-terms estimated by using LPCAL on 3C 279 between epochs, which can be attributed to the structural changes of the calibrator. 

In Table~\ref{tab:dtermstability}, we present the average $L_1$ norm for different calibration versions for the BW088A/G data sets and the BG250A/BG250B1 data sets. The D-term solutions are the most stable for the GPCAL estimates, followed by the LPCAL on M87 estimates, which is consistent with the results of our synthetic data test (Section~\ref{sec:synthetic}) and the closure trace analysis (Section~\ref{sec:ctrace}). We found that the large $L_1$ norm for the LPCAL-OJ 287 estimates for the latter epoch data sets is due to a very large deviation in the D-terms of SC antenna which comprises very long baselines. This result is also consistent with the high $\chi^2_{\mathcal{C}, {\rm 3C 279}}$ value obtained when using LPCAL on OJ 287 (see Table~\ref{tab:ctrace}). We expect the inaccurate D-terms of SC originate from its limited parallactic angle coverage of OJ 287.

\begin{deluxetable}{ccccc}
\tablecaption{$\langle L_1 \rangle$ of D-term estimates between epochs \label{tab:dtermstability}}
\tablewidth{0pt}
\tablehead{ & \colhead{GPCAL} & \colhead{LPCAL$_{\rm M87}$} & \colhead{LPCAL$_{\rm OJ287}$} & \colhead{LPCAL$_{\rm 3C279}$}}
\startdata
BW088A/BW088G & 0.44 & 0.56 & 0.82 & 0.96\\
BG250A/BG250B1 & 0.56 & 0.64 & 1.75 & 1.00 \\
\enddata
\tablecomments{Average $L_1$ norm of the D-term estimates between epochs in units of \% for different calibration versions.}
\end{deluxetable}

\section{The effect of a breakdown of the similarity approximation for calibrators}
\label{appendix:bm413i}

In this appendix, we investigate the effect of a breakdown of the similarity approximation in LPCAL for the calibrators 3C 279 and OJ 287 using VLBA data. Our synthetic data test (Section~\ref{sec:synthetic}) demonstrated that LPCAL using these calibrators cannot reconstruct D-terms accurately because they do not satisfy the similarity approximation. We can investigate this effect using real data as well if there are many sources observed in the same run. In this case, one can obtain the distribution of D-terms obtained with LPCAL using many sources. The true D-term will likely be located near the center of the distribution. This approach has been adopted for instrumental polarization calibration of the large survey observing programs \citep[e.g.,][]{Jorstad2005, Hovatta2012}. We can compare the estimates obtained with LPCAL and GPCAL with the D-term distribution to test which one gives more accurate solutions.

\begin{figure*}[t!]
\centering
\includegraphics[trim = 0mm 0mm 0mm 0mm, width = 0.49\textwidth]{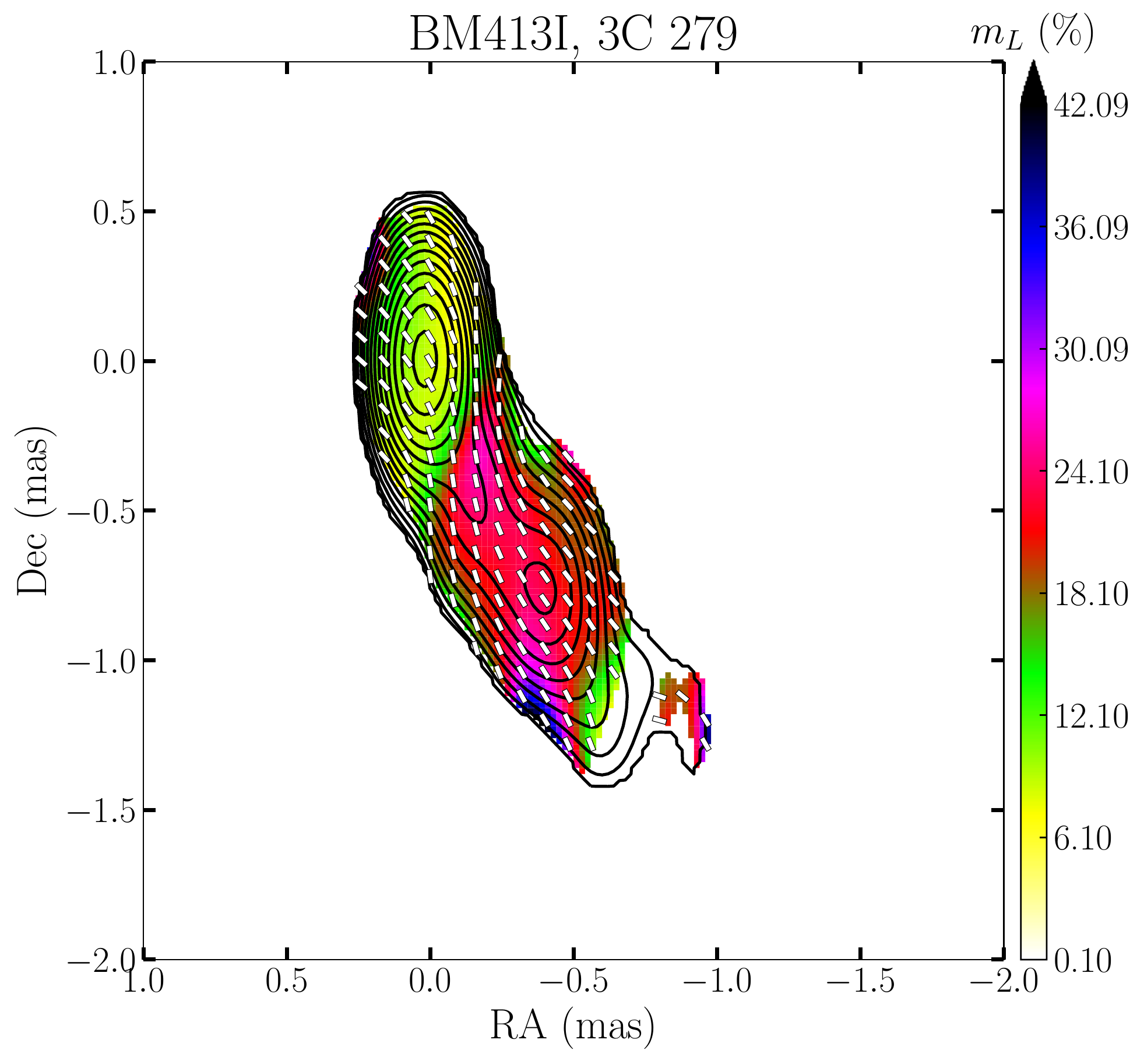}
\includegraphics[trim = 0mm 0mm 0mm 0mm, width = 0.49\textwidth]{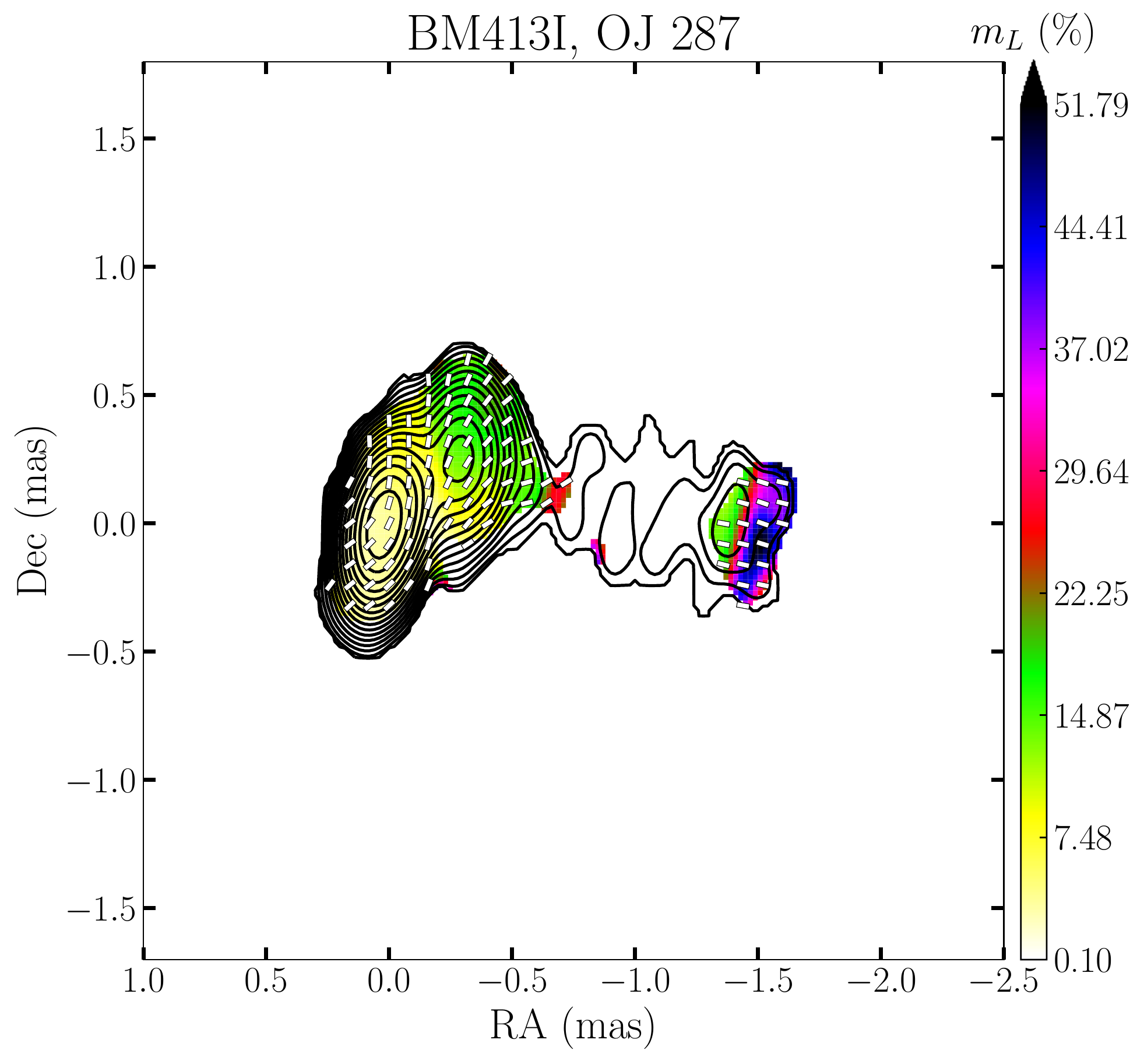}
\caption{Linear polarization images of 3C 279 (left) and OJ 287 (right) from the BM413I data. Color shows fractional polarization in units of \%; contours show total intensity; white ticks show EVPAs. \label{fig:bm413i_maps}}
\end{figure*}

We used the VLBA data observed as part of the VLBA-BU-BLAZAR monitoring program at 43 GHz\footnote{\url{https://www.bu.edu/blazars/VLBAproject.html}} \citep{Jorstad2017} on 2015 July 2 (project code: BM413I). This data was analyzed and the results were shown in \cite{Park2021a}. We ran LPCAL on 23 sources individually, and ran GPCAL using 3C 279 and OJ 287 for both initial D-term estimation using the similarity approximation and 10 iterations of instrumental polarization self-calibration. We compute the $L_1$ norm for each D-term component between the median of the LPCAL estimates and the GPCAL and individual LPCAL estimates. We found average $L_1$ norms (averaged over antennas, IFs, D-term components) of $0.62, 1.03$, and $1.34\%$ for the GPCAL, LPCAL-OJ287, and LPCAL-3C279 estimates, respectively. This result supports the conclusion derived based on the synthetic data test and the closure trace analysis that LPCAL using the calibrators cannot reconstruct accurate D-terms due to a breakdown of the similarity approximation. We present the linear polarization images of the calibrators from the BM413I data in Figure~\ref{fig:bm413i_maps} to demonstrate that the similarity approximation does not hold well for these sources. The fractional polarizations and EVPAs gradually change even within each knot-like structure in the total intensity image, making it difficult to apply the similarity approximation even if the total intensity model is divided into several submodels.

\begin{figure*}[t!]
\centering
\includegraphics[width = 0.49\textwidth]{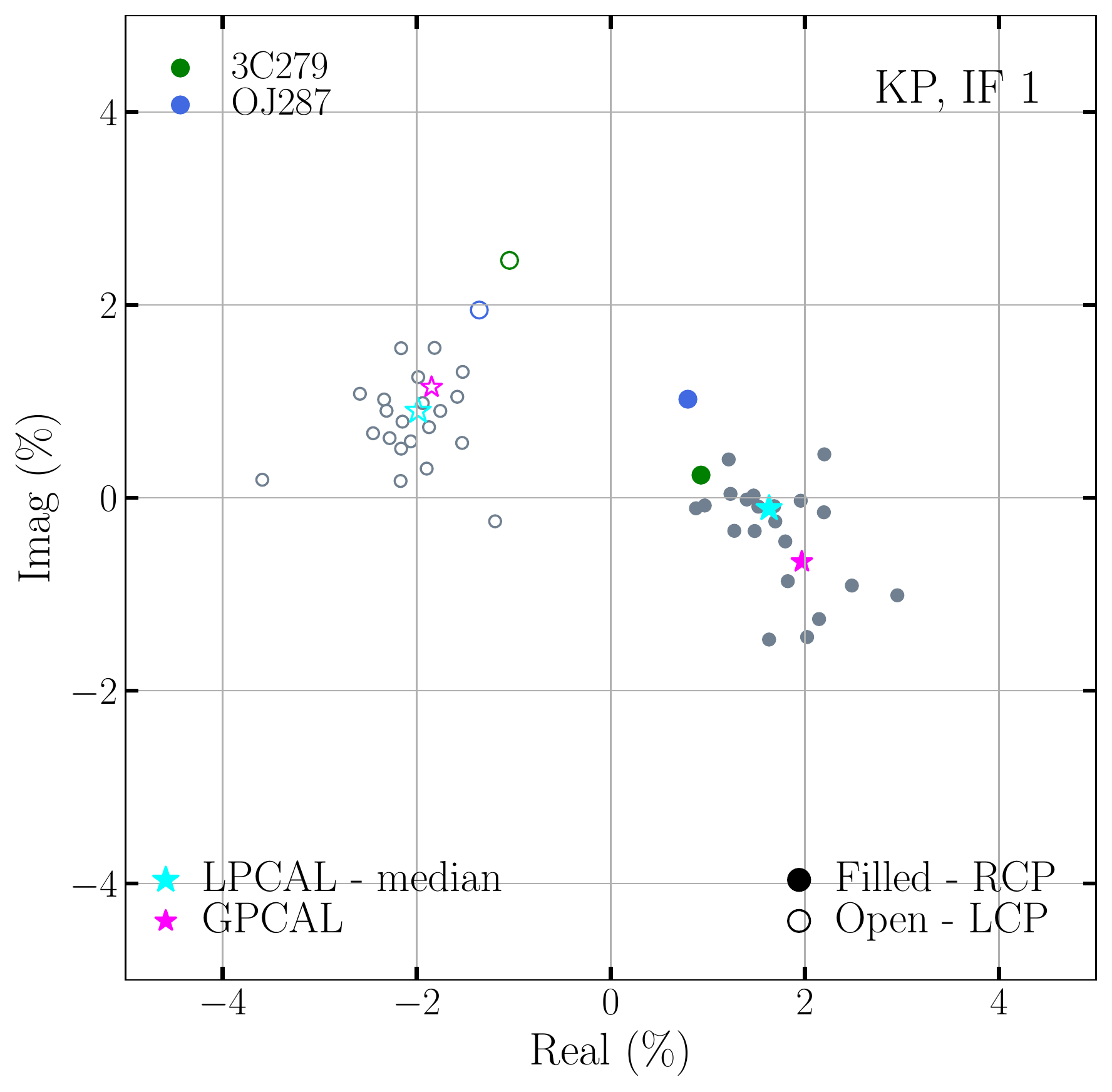}
\includegraphics[width = 0.495\textwidth]{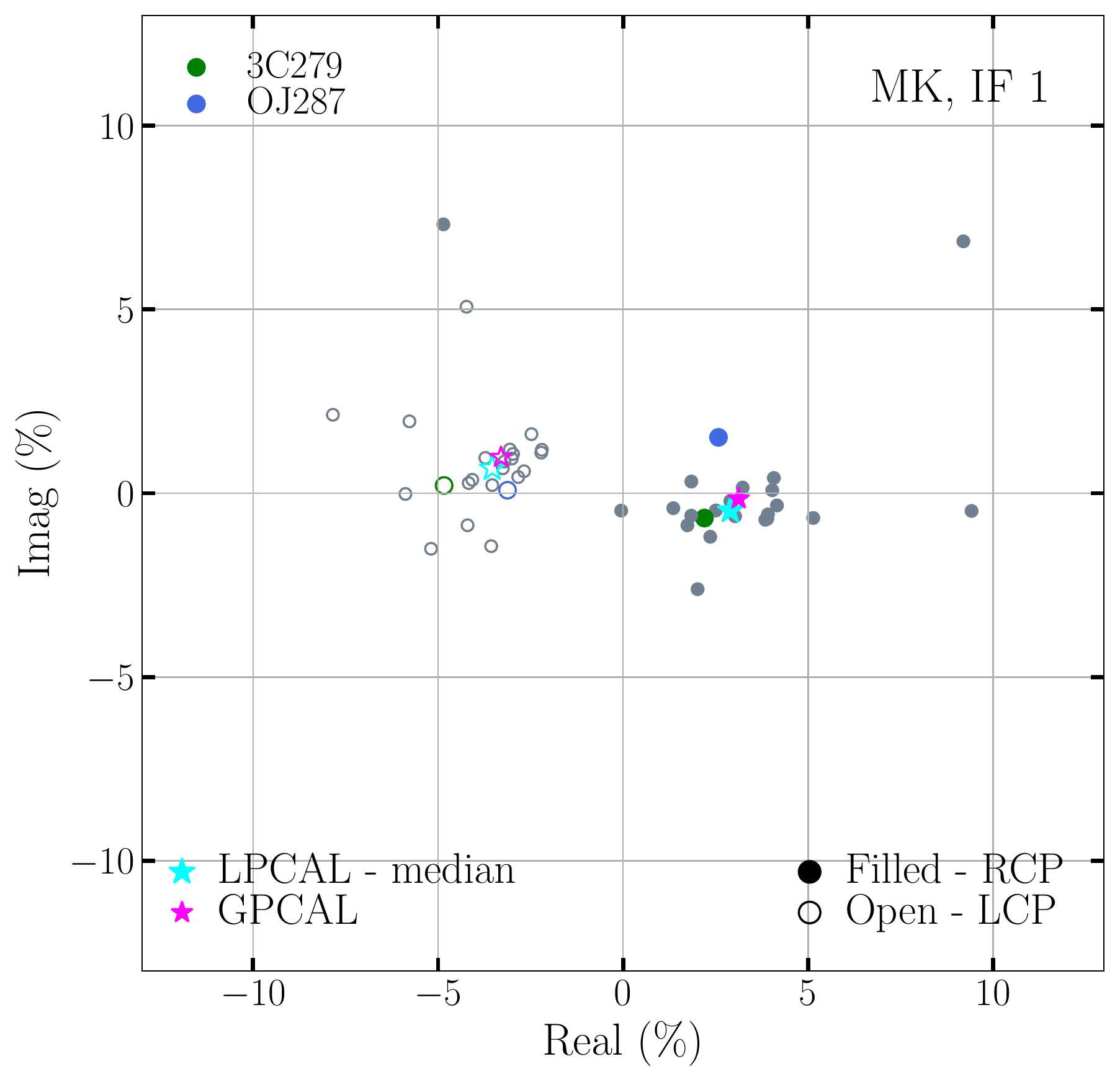}
\caption{Distribution of D-term estimates obtained with LPCAL using individual sources (circles), their median values (cyan stars), and the GPCAL estimates using 3C 279 and OJ 287 (magenta stars) for KP and MK antennas for IF 1 of the BM413I data. The filled and open symbols denote D-terms for RCP and LCP, respectively. The LPCAL estimates from 3C 279 and OJ 287 are shown in green and blue colors, respectively. \label{fig:bm413i}}
\end{figure*}

In Figure~\ref{fig:bm413i}, we present an example D-term distribution of the LPCAL estimates for individual sources and the GPCAL estimates for KP and MK antennas for IF 1, which comprise short and long baselines, respectively. The individual LPCAL estimates are scattered but concentrated on certain locations on the complex plane, represented by the median of the D-term components (cyan stars). The GPCAL estimates (magenta stars) are very close to the median of the LPCAL estimates. However, the LPCAL estimates from 3C 279 (blue dot) and OJ 287 (green dot) show larger deviation from the median values.

\bibliography{AAS33732R1.bib}{}
\bibliographystyle{aasjournal}

\end{document}